\documentclass[orsc,final]{informs3}
\OneAndAHalfSpacedXI 

\usepackage{musicography}
\usepackage{natbib}
\usepackage{mathtools}
\usepackage{changes}
\usepackage{color,soul}
\usepackage{tikz} 
\usetikzlibrary{shapes.geometric}
\usetikzlibrary{shapes.misc}
\usepackage{hyperref}
\usepackage[T1]{fontenc}
\hypersetup{hidelinks}
 \bibpunct[, ]{(}{)}{,}{a}{}{,}%
 \def\bibfont{\small}%
\TheoremsNumberedThrough     
\ECRepeatTheorems

\usepackage{epstopdf} 
\usepackage{paralist} 
\usepackage{multirow}
\usepackage{threeparttable}
\usepackage{adjustbox}

\usepackage{ctable} 
\usepackage{vcell} 
\usepackage{caption}
\usepackage{array}
\newcolumntype{L}[1]{>{\raggedright\let\newline\\\arraybackslash\hspace{0pt}}m{#1}}
\newcolumntype{C}[1]{>{\centering\let\newline\\\arraybackslash\hspace{0pt}}m{#1}}
\newcolumntype{R}[1]{>{\raggedleft\let\newline\\\arraybackslash\hspace{0pt}}m{#1}}
\usepackage{makecell}

\EquationsNumberedThrough    

\MANUSCRIPTNO{OS-MS-2023-17357.R2}

\def\ie{\emph{i.e.}}
\begin{document}

\RUNAUTHOR{Böttcher and Klingebiel}

\RUNTITLE{Organizational Selection of Innovation}
\TITLE{Organizational Selection of Innovation}

\ARTICLEAUTHORS{%
\AUTHOR{Lucas Böttcher}
\AFF{Dept.\ of Computational Science and Philosophy, Frankfurt School of Finance and Management, 60322 Frankfurt, Germany, and Dept.\ of Medicine, University of Florida, Gainesville 32610, FL, USA, \EMAIL{l.boettcher@fs.de}}
\AUTHOR{Ronald Klingebiel}
\AFF{Frankfurt School of Finance and Management, 60322 Frankfurt, Germany}
}

\ABSTRACT{Budgetary constraints force organizations to pursue only a subset of possible innovation projects. Identifying which subset is most promising is an error-prone exercise, and involving multiple decision makers may be prudent. This raises the question of how to most effectively aggregate their collective nous. Our model of organizational portfolio selection provides some first answers. We show that portfolio performance can vary widely. Delegating evaluation makes sense when organizations employ the relevant experts and can assign projects to them. In most other settings, aggregating the impressions of multiple agents leads to better performance than delegation. In particular, letting agents rank projects often outperforms alternative aggregation rules \textemdash~including averaging agents' project scores as well as counting their approval votes \textemdash~especially when organizations have tight budgets and can select only a few project alternatives out of many.
}%


\maketitle
%
\section{Introduction}
We examine rules for aggregating individual project evaluations so as to make organizational portfolio selection decisions. Such resource-allocation decisions have specific characteristics \citep{klingebiel2018risk,levinthal2017resource,sengul2019allocation}. They are made intermittently, with organizations considering multiple resource-allocation alternatives at a time. They are subject to budget constraints, limiting the number of alternatives an organization can pursue. And they are made under uncertainty, exposing organizations to allocation errors.

Consider how organizations select from a slate of innovation ideas \citep{brasil2019product, kavadias2020framework}. A group of executives with different backgrounds meet periodically to review funding proposals. The resources they are permitted to allocate suffice for only a fraction of the proposals before them. Many of the proposals are outside of executives' prior experience, leading to noisy assessments. They may thus consider delegating the portfolio-selection decisions to the relatively most qualified person, or to combine their limited expertise in various ways to arrive at better decisions. Which approach to arriving at an organizational portfolio will yield the best results in expectation?

The study of organizational aggregation processes is a resurgent scholarly pursuit \citep{knudsen2007two,puranam2018book}. Of particular relevance for our study is the finding that aggregating project approval decisions through majority voting usually leads to better outcomes than attainable through averaging scores \citep{csaszar2013organizational}. Delegation is effective when relevant executive expertise is available and evident, or when organizations struggle to afford coordinating among a wider set of decision makers \citep{healey2023costs}. 

We advance such insights into aggregation by studying portfolio selection. Choosing a subset of available project alternatives is different from project approval in two ways. First, portfolio selection requires decision makers to observe a budget constraint. Second, to identify the best subset of projects to be funded, portfolio selection involves discrimination. 

One implication of these unique features is a different performance dimension. What matters for portfolio selection is maximizing the expected performance of the projects chosen for funding, not the performance of every project approvable in isolation. Many of the latter may in fact not make the cut. Another implication is that portfolio selection, unlike isolated project approval, involves prioritization. Rank-order approaches discussed in the social-choice literature on multi-winner voting \citep{elkind2017properties,pacuit2019voting} thus become relevant. 

The mathematical model of portfolio selection we develop contains heterogeneously informed, non-strategic agents who are given the task of selecting from a list of independent project proposals. Different rules for aggregating agents' selection decisions produce variation in performance. As our model is intended as a first step for studying organizational decision making at the portfolio level, we leave potentially interesting richness such as project interdependence, decision-maker interaction, and strategic behavior to future research.

We find that relying on individuals is almost always inferior to aggregating multiple opinions. Majority voting performs poorly when resource constraints require organizations to be highly selective. Averaging performs better but is often outdone by a simple process of having agents produce a ranked preference list. Totaling such ranks is the most performative method of aggregation in many scenarios, inferior to delegation only when firms know that they have the right experts for evaluation.

The dominance of Ranking \textemdash~based on an aggregation process known as Borda count \citep{brandt2016handbook} \textemdash~is due to its robustness against project-quality misclassification that degrades the performance of other selection methods like Averaging more substantially. Organizational selection of innovation thus benefits from a relatively crude ordering process that differs from the voting procedures prior work would recommend for the isolated approval of individual projects.

Our work provides insights into how organizations can harness collective decision-making processes to effectively allocate resources. In the attempt to understand the management of innovation portfolios \textemdash~of patents, drug candidates, or technology ventures~\citep{eklund2022pharma,kumar2023patents,toh2022integration}, for example \textemdash~empirical work honed in on group-decision biases such as those concerning novelty \citep{criscuolo2017evaluating} or commitment \citep{klingebielesser2020}. Gauging the meaningfulness of selection biases requires outlining the performance that can be achieved in the absence of bias. Our work provides such expectations for multiple selection procedures. It offers a structured answer to organizations searching for and experimenting with different aggregation methods \citep{carson2023grants,luo2021moviescripts,sharapovdahlander2021}. Our work thus opens up avenues for future research on the performance of decision-making structures, particularly as regards rules for aggregating selection under uncertainty.
\section{Innovation Portfolio Selection}
The starting point of our work is the model of \cite{csaszar2013organizational}. They compare the performance of collective decisions \textemdash~voting and averaging \textemdash~with that of individuals \textemdash~anyone and experts. For detail on the research preceding the Csaszar and Eggers model, we refer to the authors' comprehensive review of the field. Since then much work, mostly non-organizational, has focused on how weighted algorithms can improve crowd wisdom \citep{budescu2024crowds,kameda2022crowds,xia2022groupai}. Csaszar and Eggers' work remains the most relevant baseline for our purposes, because it considers the organizational context of projects with uncertain payoffs\footnote{\cite{oraiopoulos2020diversecommittees} model the isolated approval of uncertain projects. They examine the effect of preference diversity on the performance of majority voting. In our model, we account for more aggregation rules, but exclude strategic behavior.} and variously informed decision makers, central features of organizational reality and part of what motivates our research.

We extend the Csaszar and Eggers model to the organizational selection of multiple project candidates, subject to resource constraints. Concurrent consideration is common when organizations or investors review a list of innovative proposals and select only those alternatives they deem most worthy of receiving resources from limited budgets. They rarely approve proposals without considering funding limits and opportunity costs \citep{cooper1993screening,wheelwright1992npd}, since each dollar spent is one not spent elsewhere.

Organizations instead aim to make the most of the few resources at their disposal.\footnote{Firms may want to maximize returns at some level of risk. For example, financial portfolios often contain assets with potentially sub-optimal return expectations to diversify sources of risk. Our present work, however, does not require the additional consideration of hedging goals. The payoffs from projects in our model are independent of each other and none is structurally more at risk than others. Relaxing these constraints would require arbitration among multiple goals \citep{faliszewski2017dualgoals}, a phenomenon worthy of further empirical research on preferences.} Therefore, for projects to be selected into the portfolio, they need to not only clear a quality threshold such as an expected rate of return, but also be of higher quality than concurrently reviewed alternatives. Discrimination among projects not only complicates the application of the decision rules discussed in prior work. It also gives rise to an additional class of rules involving relative preferences that Csaszar and Eggers did not have to consider. This departure likely affects which rule helps organizations perform best.

Relative preferences and the general problem of selecting a subset of alternatives feature in the social-choice literatures on multi-winner voting systems \citep{nitzan2017juries} and participatory budgeting \citep{benade2021preference,goel2019knapsack}. A primary subject of inquiry in such social-choice research is how closely collective decisions reflect individual preferences, but a notable sub-stream additionally examines how collectives reach \textquotedbl correct\textquotedbl~decisions \citep[e.g.][]{austen1996rationality,nitzan1982optimal}. 

Identifying the single best option in a set of more than two uncertain alternatives is a task in which majority voting still performs well with rules for tie-breaking \citep{hastie2005robust,young1988condorcet}. We extend this insight by asking which aggregation method should be used if organizations want to select multiple projects \textemdash~ the best subset of opportunities that the organizational budget allows them to afford. Here, the multi-winner literature already foreshadows the usefulness of ranking methods \citep{procaccia2012mle}. Large sets of homogeneous voters identify the correct order of noisily perceived choices more often when ranking, rather than approving, the choices \citep{boehmer2023bias, faliszewski2017dualgoals, rey2023conflict}. 

Generating similar insights for portfolio decision rules matters. Organizations have few decision makers, and with heterogeneous expertise. Aggregating their impressions is a problem that receives attention: Firms have been found to engage in costly trial-and-error search for innovation-selection processes \citep{sharapovdahlander2021}. Some venture capitalists deliberately adopt minority voting \citep{malenko2023vc}, for example, in the attempt to improve performance by reducing errors of omissions in environments where success follows a power-law distribution. Broadly speaking, however, empirical work in this area suggests that firms are not particularly effective in making selection decisions \citep{criscuolo2017evaluating,klingebiel2015regimes,sommer2020search}. Our work thus aims to establish conditions under which one can expect different forms of aggregation to improve the performance of organizational portfolio selection.
\section{Modeling Portfolio Selection}
\label{sec:model_setup}
Portfolio selection occurs whenever multiple candidates vie for limited resources. While one can easily imagine a court case to be judged in isolation, with culpability determined irrespective of the outcomes from other concurrent cases, it is harder to imagine companies to decide funding for an innovation project irrespective of superior alternatives. Organizations will want to spend scarce funds on innovation projects only if they perceive future payoffs to be in excess of those of other projects. Even when organizations proceed with a single project only, the decision likely resulted from a process of selection, rather than an isolated instance of project assessment \citep{si2022managing}. 

\begin{figure}[htbp]
     \FIGURE
    {\includegraphics[width=0.7\textwidth]{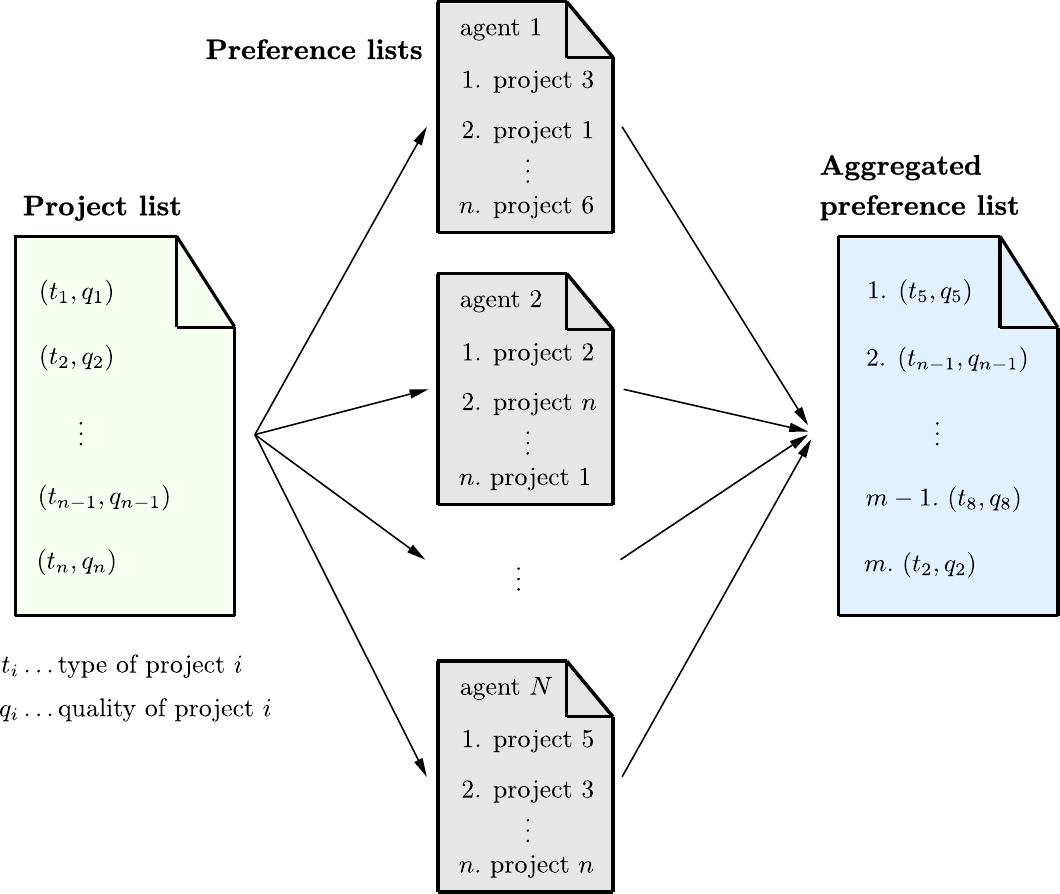}}
    {Organizational Portfolio Selection\label{fig:schematic}}
    {$N$ agents consider a list of $n$ projects of types $t_i$ and qualities $q_i$ $(i\in\{1,\dots, n\})$. Agents $j\in \{1,\dots, N\}$ compile preference lists that are ordered on the basis of their project-quality perceptions $q_{ij}'$. Finally, an aggregation rule combines the individual preference lists.}
\end{figure}

Therefore, we introduce selection into the organizational decision framework of \cite{csaszar2013organizational} by adding a budget constraint of $m\geq 1$  projects, chosen from $n\geq m$ alternatives. Agents' evaluations of projects inform an organization's selection of a subset (see Figure~\ref{fig:schematic}).

We consider both $m$ and $n$ exogenous. In established organizations, top management sets aside a portion of organizational resources for innovation. Top management determines this budget $m$ by gauging the need for rejuvenation and considering rival demands for the same resources \citep{schilling2023book}. Innovation executives, who are to be our model agents, then decide on how to spend the given budget. In real-world organizations, innovation executives might influence the budget-setting process and occasionally request increases alongside emerging opportunities \citep{klingebiel2013unkunk}. We leave the examination of such exceptions to future research. 

Likewise, we treat $n$ as independent from our agents. The project candidates reviewed at an innovation-board meeting are typically generated by personnel other than the decision makers \citep{wheelwright1992npd}. The possibility that innovation executives are partial to the generation and evaluation of some but not other opportunities \citep{dushnitsky2023randomisation}, or that they may revisit initial decisions at later points \citep{furr2021renewal}, are extension areas for future research.

Following \cite{csaszar2013organizational}, we characterize project candidates with two stochastic variables: $t_i\sim\psi$ represents the type, and $q_i\sim\phi$ the quality of project $i\in\{1,\dots,n\}$. The distributions $\psi$ and $\phi$ have supports $[\underline{t},\overline{t}]$ and $[\underline{q},\overline{q}]$, respectively (see Table~\ref{tab:model_variables}). 

\begin{table}[htbp]
\TABLE
{Model Components\label{tab:model_variables}}
{\centering
\renewcommand*{\arraystretch}{1.6}
\small
\begin{tabular}{| >{\centering\arraybackslash} m{13em}| 
>{\raggedright\arraybackslash} m{25em}|}\hline
\multicolumn{1}{|c|}{\textbf{Symbol}} & \multicolumn{1}{c|}{\textbf{Definition}}
\\[1pt] \hline\hline
 \,\,\, $n\in\{1,2,\dots\}$\,\, & number of available projects \\[1pt]  \hline
 \,\,\, $m\in\{1,2,\dots,n\}$\,\, & number of selected projects \\[1pt]  \hline
 \,\,\, $N\in\{1,2,\dots\}$\,\, & number of agents \\[1pt]  \hline
\,\,\, $q_i \in [\underline{q},\overline{q}]$\,\, & quality of project $i$ \\[1pt]  \hline
 \,\,\, $t_i\in [\underline{t},\overline{t}]$\,\, & type of project $i$ \\[1pt]  \hline
 \,\,\, $q_{ij}' \in \mathbb{R}$\,\, & quality of project $i$ as perceived by agent $j$ \\[1pt]  \hline
 \,\,\, $e_{j}\in[\underline{e},\overline{e}] $\,\, & expertise value of agent $j$ \\[1pt]  \hline
  \,\,\, $\eta_{ij}\in[\underline{\eta},\overline{\eta}] $\,\, & perceptual noise of agent $j$ w.r.t.\ project $i$  \\[1pt]  \hline
 \,\,\, $\phi$\,\, & distribution of project qualities $q_i$ with support $ [\underline{q},\overline{q}]$ \\[1pt]  \hline
 \,\,\, $\psi$\,\, & distribution of project types $t_i$ with support $ [\underline{t},\overline{t}]$ \\[1pt]  \hline
 \,\,\, $\chi$\,\, & distribution of expertise values $e_{j}$ with support $[\underline{e},\overline{e}]$ \\[1pt]  \hline
\end{tabular}
\vspace{1mm}}
{An overview of the main model variables and distributions.}
\end{table}

Project type $t_i$ can be viewed as a variable describing knowledge domains. Incorporating such domains means that agents cannot assess all projects equally well, a departure from social-choice models of multi-winner elections \citep[e.g.][]{procaccia2012mle}.

Agents $j\in\{1,\dots,N\}$ are characterized by expertise values $e_{j}$, which are distributed according to a distribution function $\chi$ with support $[\underline{e},\overline{e}]$. Accordingly, $q_{i j}'=q_i+\eta_{ij}$ denotes the quality of project $i$ as perceived by agent $j$, where $\eta_{ij}$ is distributed according to a normal distribution $\mathcal{N}(0,|t_i-e_{j}|)$ with zero mean and standard deviation $|t_i-e_{j}|$. The inclusion of a noise term accounts for uncertainty in project evaluation. Noise thus varies with domain expertise; the quantity $|t_i-e_{j}|$ captures the degree to which project type matches agent expertise. 

This operationalization of variation in judgement quality is a plausible approximation of the organizational reality in portfolio decision making under uncertainty \citep{kornish2017research}. It endows agents with equally imperfect capabilities but recognizes that they may come from different backgrounds. For example, the innovation boards at pharmaceutical companies encompass experts from different therapeutic classes \citep{aaltonen2020schering}. Those experts' judgements are more accurate for proposals in their own class. Domain-specific expertise has been documented to similarly influence innovation decision quality at device manufacturers \citep{vinokurova2023kodak} and service firms \citep{klingebielesser2020}. We thus follow \cite{csaszar2013organizational} in recognizing this feature of evaluative precision in our model.\footnote{Our design choice of domain-specific expertise mirrors Hotelling models, in which actors have different distances to a focal point~\citep{hotelling1929,NOVSHEK1980313}. We refer the reader to \cite{adner2014positioning} for a corresponding review. Alternatives to the Hotelling approach include belief-updating models, in which decision makers share identical priors but receive different signals \citep[][]{li2001aercommittees,oraiopoulos2020diversecommittees} that together result in project assessments. This approach produces judgment-specific, rather than expert-specific, variation in precision \citep{einhorn1977quality,hogarth1978note}. Alternatively, one could conceive of expertise as a vector \citep{csaszar2016mental}. For instance, one dimension of expertise may pertain to environmental sustainability aspects and another to mechanical design aspects of project value. Multi-dimensional representations might reflect the micro-foundations of the decision-making challenge in more detail \textemdash~yet we do not expect the replacement of the Hotelling expedience with greater knowledge dimensionality to materially affect aggregation rules' efficacy in dealing with judgment imprecision. We would welcome future research on this topic.}

Building on previous work on multi-winner electoral systems~\citep{brandt2016handbook,elkind2017properties}, we represent the quantities used to produce aggregated preference lists by an ordered triplet $M=(Q,T,E)$, where $Q=\{q_1,\dots,q_n\}$, $T=\{t_1,\dots,t_n\}$, and $E=\{e_{1},\dots,e_{N}\}$ denote one realization of the sets of project qualities, project types, and expertise values, respectively. Each agent sorts $n$ projects in descending order of perceived quality. For example, in the case of $n=2$ available projects, agent $j$ strictly prefers the first over the second project if her perception of project qualities satisfy $q_{1j}'> q_{2j}'$. In general, the relation $i \succ_j k$ means that agent $j$ strictly prefers project $i$ over $k$, which is the case if and only if $q_{ij}'>q_{kj}'$. To denote the position of project $i$ in the preference order of agent $j$, we use the notation ${\rm pos}_{j}(i)$.

An aggregation rule $\mathcal{R}(M,m)$ is a function that maps a realization of $M=(Q,T,E)$, to a corresponding subset of $m\leq n$ selected projects. Ties occur if the selection rule produces multiple outcomes of the same cardinality. If ties occur in our simulations, we uniformly at random select one outcome. 

We use $f_{q_{(i)}}^{(\mathcal{R})}$ to denote the PDF of $q_{(i)}^{(\mathcal{R})}$, the ordered project qualities under selection rule $\mathcal{R}$. The support of $f_{q_{(i)}}^{(\mathcal{R})}$ is $[\underline{q},\overline{q}]$. The expected portfolio performance associated with $q_{(i)}^{(\mathcal{R})}$ under selection rule $\mathcal{R}$ thus is

\begin{equation}
\mathbb{E}^{(\mathcal{R})}[q;m,n] = \sum_{i=n+1-m}^n \int_{\underline{q}}^{\overline{q}} q\, f_{q_{(i)}}^{(\mathcal{R})}(q)\,\mathrm{d}q\,.
\label{eq:performance_R}
\end{equation}

Dividing this by $m$ would yield the corresponding expected quality per selected project. Analytic expressions of $f_{q_{(i)}}^{(\mathcal{R})}$ and analytically evaluating the integral in Eq.~\eqref{eq:performance_R} are not tractable for general selection rules $\mathcal{R}$. Hence, our main approach involves running Monte Carlo simulations of independent and identically distributed (i.i.d.) realizations of various selection rules. 

Appendix~\ref{app:sensitivity} derives the bounds within which we can expect portfolio performance to vary in the simulations. For a uniform quality distribution $\phi\sim\mathcal{U}(\underline{q},\overline{q})$, the theoretical maximum of the expected portfolio performance is
\begin{equation}
\mathbb{E}^*[q;m,n] =m \left[\underline{q} +(\overline{q}-\underline{q})\frac{2n+1-m}{2n+2}\right]\,
\label{eq:uniform_e_max}
\end{equation}
It constitutes an upper limit against which we can evaluate different selections rules $\mathcal{R}$.
\section{Aggregation Rules}
\label{sec:selection_rules}
The aggregation rules $\mathcal{R}$ that we adapt and examine in our Monte Carlo simulations encompass classics that are simple and distinctive, a subset of potentially endless method variants \citep{elkind2017properties}. They encompass the voting and scoring rules considered in \cite{csaszar2013organizational} plus a simple ranking rule known in the social-choice literature as Borda count \citep{elkind2017properties}.

All our rules preserve the balance of type I and type II errors in expectation \citep{klingebiel2018risk} and so disregard methods involving consensus, sequencing, or hierarchies that would skew the balance~\citep{sah1988committees,christensen2021context, malenko2023vc}. We also assume non-strategic agents (in contrast to \cite{piezunka2023} or \cite{marino2005decision}, for example) as well as independent projects, disregarding potential benefits from composing portfolios with projects of varying novelty or knapsack efficiency~\citep{faliszewski2017dualgoals,si2022managing}. Our decision makers neither communicate nor learn from one another or across projects \citep[see e.g.][]{becker2022thresholds,elhorst2021contest,flache2017models}. Relaxing some of these constraints would be a natural next step for considering additional aggregation rules in future research.

To transport classic aggregation rules to a portfolio-selection context, we modify them such that they impose a funding constraint at the organizational level. Selection criteria, therefore, are not based on thresholds, such as a positive average evaluation, or a majority of yes votes, that one would find in the context of isolated project approvals. Instead, organizations select into the portfolio $m$ projects with the relatively highest scores.\footnote{Our selection rules could additionally impose a project-quality threshold. For example, few executives would suggest committing to projects that they expect to yield negative payoffs. Because the parameterization of our main model effectively prevents such projects to be among the top $m$ (see Section \ref{sec:base_case} and Appendix \ref{app:distributions}), we chose to minimize rule complexity. Future work may adopt hurdle rates as required.} The subsequent definitions thus incorporate organizational discrimination.

\paragraph{Individual.} All projects are evaluated by a single agent with expertise value $e_{\rm M}$, which is the mean of the expertise distribution. The organization then ranks projects based on the agent's quality perceptions and selects the top $m\in\{1,\dots,n\}$ projects. This selection rule implements the Individual rule of \cite{csaszar2013organizational} in a portfolio context.

\paragraph{Delegation.} Each project is evaluated by the agent whose expertise is most closely aligned with the project's type. These are agents whose expertise value $e_{j}$ minimizes the uncertainty $|t_i-e_{j}|$). The organization then ranks projects based on experts' quality perceptions and selects the top $m\in\{1,\dots,n\}$ projects. This selection rule implements the Delegation rule of \cite{csaszar2013organizational} in a portfolio context.\footnote{Instead of delegating projects to different experts, organizations may consider assigning the responsibility to a single portfolio expert. In this scenario, all projects would be assessed by the agent whose expertise minimizes the overall uncertainty $\sum_{i=1}^n |t_i-e_{j}|$. The organization then ranks projects based on the expert's quality perceptions and selects the top $m\in\{1,\dots,n\}$ projects. The expertise that minimizes the overall uncertainty is equal to the mean type. The approach is thus equivalent in expectation to the Individual rule we state above.}

\paragraph{Voting.} All projects are evaluated by all agents. Agents allocate a vote to each project for which have a positive perception of quality. The organization then ranks projects based on the number of agent votes and selects the top $m\in\{1,\dots,n\}$ projects. This selection rule implements in a portfolio context the Voting rule used by \cite{csaszar2013organizational}\footnote{Note that the model of \cite{csaszar2013organizational} is not multi-candidate voting, since decision makers never consider projects concurrently. They rather (dis)approve each project in isolation. The model of \cite{csaszar2018crowdpops}, with crowds voting for one of two projects, is closer to a multi-candidate setting.} and others \citep[e.g.][]{keuschnigg2017abilities,li2001aercommittees,oraiopoulos2020diversecommittees}.

\paragraph{Averaging.} All projects are evaluated by all agents. The organization then ranks projects based on agents' mean quality perceptions \textemdash~scores, effectively \textemdash~and selects the top $m\in\{1,\dots,n\}$ projects. This selection rule implements the Averaging rule of \cite{csaszar2013organizational} in a portfolio context.

\paragraph{Ranking.} All projects are evaluated by all agents. Each agent $j$ places the projects in a descending order of perceived quality. Each project $i$ thus receives a position ${\rm pos}_{j}(i)$. The organization then ranks projects based on the sum of agents' reversed project positions, ${n-{\rm pos}_{j}(i)}$, and selects the top $m\in\{1,\dots,n\}$ projects. This selection rule implements the Borda rule of the social-choice literature ~\citep{elkind2017properties} in a portfolio context.\footnote{Although not in the realm of uncertain innovation projects, a public example application of the Ranking rule is the Aggregate Ranking of Top Universities (\url{https://research.unsw.edu.au/artu/methodology}). In it, the University of New South Wales aggregates the preference lists of three agents: Times Higher Education, Quacquarelli Symonds, and ShanghaiRanking Consultancy. They each form their quality perceptions for hundreds of universities based on a list of different criteria. The rank that agents assign to a university then automatically results from these scores. The Aggregate Ranking of Top Universities could be used to create a portfolio of $m$ best universities.} 
\section{Results}
Our base-case analyses use the parameter values of \cite{csaszar2013organizational} to enable comparisons. The number of decision makers is set to $N=3$, the type distribution to $\psi=\mathcal{U}(0,10)$, the quality distribution to $\phi=\mathcal{U}(-5,5)$, and the noise distribution to $\varphi=\mathcal{N}(0,|t_i-e_{j}|)$. We additionally set the number of available projects to $n=100$.

The expertise of the agent in the Individual rule is set to a central $e_{\rm M}=5$. To represent the collective knowledge of an organization's decision makers, we assign each agent $j\in\{1,\dots,N\}$ an expertise value $e_j=e_{\rm M}-\beta+\frac{2\beta}{N-1} (j-1)$, where $\beta\in[0,5]$ denotes the knowledge breadth of an organization.

For the given distributions, we generate i.i.d.\ realizations of the underlying model quantities to perform Monte Carlo simulations of all aggregation rules presented in Section~\ref{sec:selection_rules}. We then compare their portfolio performances, $\mathbb{E}^{(\mathcal{R})}[q;m,n]$, and explore variation in the parameter space to probe for generality of our results.

All implementation details are provided at \url{https://gitlab.com/ComputationalScience/multiwinner-selection}.
\subsection{Aggregation-Rule Performance}
\label{sec:base_case}
\begin{figure}[htbp]
    \FIGURE
    {\includegraphics[width=\textwidth]{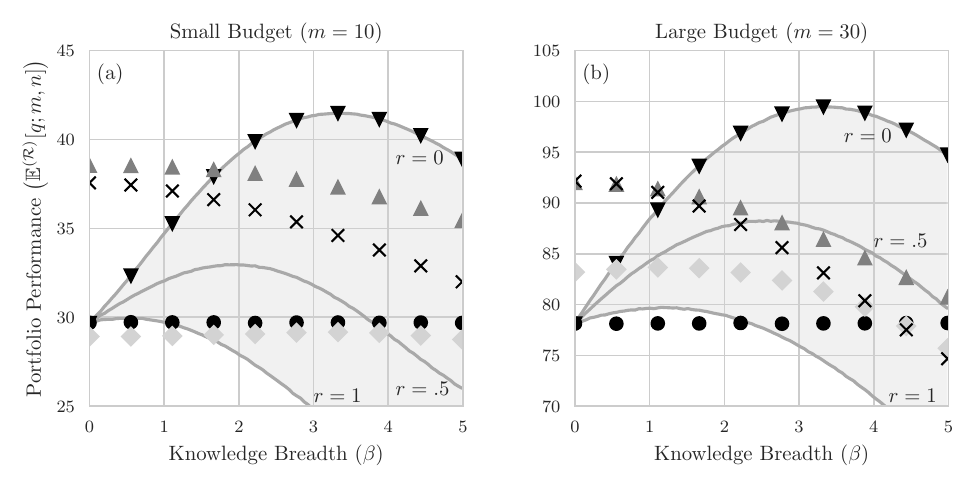}}
    {Aggregation-Rule Performance\label{fig:selection_rules}}
    {The panels show the total performance of project portfolios selected by the aggregation rules of Individual (\tikz\node[black,fill=black,circle,inner sep=2pt,draw] {};), Delegation (\tikz\node[black,fill=black,isosceles triangle,isosceles triangle apex angle=60,rotate=270,inner sep=1.8pt,draw] {};), Voting (\tikz\node[lightgray,fill=lightgray,diamond,inner sep=2pt,draw] {};), Averaging (\tikz\node[black,fill=black,cross out,inner sep=2pt,line width=1pt,draw] {};), and Ranking (\tikz\node[gray!80,fill=gray!80,isosceles triangle,isosceles triangle apex angle=60,rotate=90,inner sep=1.8pt,draw] {};). The shaded area delineates the performance range for the Delegation rule as determined by delegation-error parameter $r$ defined in Section~\ref{sec:delegation_error} below. Results are based on $n=100$ available projects. The numbers of selected projects is $m=10$ in Panel (a) and $m=30$ in Panel (b). The expertise of Individual agents is $e_{\rm M}=5$. In all other rules, we consider $N=3$ agents, with expertise of agent $j\in\{1,\dots,N\}$ being $e_j=e_{\rm M}-\beta+\frac{2\beta}{N-1} (j-1)$. The project type and quality distributions are $\mathcal{U}(0,10)$ and $\mathcal{U}(-5,5)$, respectively. All results are based on ensemble means of $2\times 10^5$ i.i.d.\ realizations. The theoretical maxima for the expected quality [see Eq.~\eqref{eq:uniform_e_max}] are approximately 44.6 ($m=10$) and 104.0 ($m=30$).}
\end{figure}

In the base case, Ranking provides the highest performing aggregation rule for $\beta\lesssim 2$ (see Figure \ref{fig:selection_rules}). Averaging approaches the performance of Ranking for smaller values of $\beta$ as the number of selected projects, $m$, increases. 

Delegation to project experts is the most effective selection protocol for $\beta\gtrsim 2$ (see Figure~\ref{fig:selection_rules}). In our portfolio-selection setting, the knowledge breadth $\beta$ at which a Delegation protocol begins to outperform Ranking is larger than the reported value of $\beta$ at which Delegation outperforms other protocols in a project approval setting~\citep{csaszar2013organizational}, at least for small budgets (\ie, small values of $m$). This observation is insensitive to project-type variations as we elaborate in Section~\ref{sec:type_variation}. 

The Delegation rule comes close to the maximum possible performance $\mathbb{E}^*[q;m,n]=44.6$ ($m=10$) and 104.0 ($m=30$) for intermediate levels of knowledge breadth $\beta$. For $\beta=0$, all decision makers have the expertise $e_{\rm M}$ of Individual decision makers. As $\beta$ increases, the expertise values of all decision makers cover a broader range of project types. Hence, decision makers with expertise values close to specific project types can be selected for intermediate values of $\beta$. If $\beta$ is too large, the distance between available and required expertise grows. 

For a general uniform type distribution $\psi\sim\mathcal{U}(\underline{t},\overline{t})$, the maximum performance of the Delegation protocol is achieved if the $N$ decision makers have expertise values
\begin{equation}
e_{j}^*=\underline{t}+(2j-1)\frac{\overline{t}-\underline{t}}{2N}\quad (j\in\{1,\dots,N\})\,.
\label{eq:delegation_max}
\end{equation}
For $N=3$ decision makers with $\psi\sim\mathcal{U}(0,10)$, the maximum performance of Delegation is thus realized at expertise values
\begin{equation}
e_1^*=\frac{5}{3}\quad e_2^*=5\quad e_3^*=\frac{25}{3}\,,
\end{equation}
that is, for $\beta=10/3\approx 3.33$ (see Figure~\ref{fig:selection_rules}).

In contrast to project approval, Voting is not very effective in portfolio selection. To see why, consider that it aggregates binary signals only. The aggregate scale for totalling the votes of $N=3$ decision makers contains four levels only. Voting thus often fails to discriminate between many projects.

To gauge the discrimination limitation of the Voting protocol, we conduct additional simulations with $2\times 10^5$ i.i.d.\ realizations. Among $n=100$ projects, 31 on average receive a full three votes from $N=3$ decision makers with knowledge breadth $\beta=0$. Analogously 28 projects receive three votes with $\beta=2.5$, and 23 projects with $\beta=5$. Therefore, with $m=10$, the Voting rule, typically selects only projects receiving a full three votes. The quality difference between the best and the 20th best project can be large but aggregate votes tend not to reveal this.\footnote{The discrimination limitation of Voting might be partially remedied by asking agents to approve $m$ projects only. With small budgets and large choice sets, such $m$-approval voting \citep{elkind2017properties} limits the number of projects that are sanctioned by all agents, providing more discrimination in the top section of the aggregated preference list of projects, and less at the bottom.} Voting thus underperforms more discriminating rules such as Ranking. With greater budgets such as $m=30$, Voting does relatively better. 

Greater knowledge breadth decreases the performance of Voting but less so than that of other aggregation rules. Therefore, as $\beta$ and $m$ increase, Averaging and Voting can achieve similar performance (see Figure~\ref{fig:selection_rules}b). In such situations, Voting caps the influence of erroneous classifications made by single agents with a unsuitable expertise. Averaging suffers relatively more quickly from the aggregation of erroneous estimates provided by agents with unsuitable expertise.

These results remain stable even with extremely small budgets that permit the selection of $m=1$ project only (see Appendix \ref{app:relative_ordering}). 
\subsection{Discrimination Effectiveness}
Why does Ranking outperform Averaging? For some intuition, consider $N=3$ agents, $n=3$ available projects, and knowledge breadth $\beta=0$. In one realization of the agents' quality perceptions $q_{ij}'$ (see Table~\ref{tab:ranking_example}) we have the preference orders: $1\succ_1 3 \succ_1 2$, $2\succ_2 3 \succ_2 1$, and $1\succ_3 2 \succ_3 3$. The organization would select project \#1 first, as its sum of reversed project positions is $4$. Project \#2 is second-most attractive, with a sum of $3$. Project \#3 would be least attractive, with a sum of $2$. 

\begin{table}[htbp]
\renewcommand*{\arraystretch}{1}
\small
	\begin{center}
 \setlength\extrarowheight{2pt}
 		\caption{Aggregation Example\label{tab:ranking_example}}
    \vspace{0.3cm}
		\begin{adjustbox}{max width=0.9\textwidth}
			\begin{threeparttable}
				\begin{tabular}{  L{0.001\textwidth}L{0.12\textwidth}C{0.06\textwidth}C{0.06\textwidth}C{0.06\textwidth}C{0.06\textwidth}C{0.06\textwidth}C{0.06\textwidth}C{0.02\textwidth}C{0.12\textwidth}C{0.12\textwidth}}
&& \multicolumn{6}{c}{Individual} & & \multicolumn{2}{c}{Organization} \\
&& \multicolumn{2}{c}{Agent $1$} & \multicolumn{2}{c}{Agent $2$} & \multicolumn{2}{c}{Agent $3$} & & \multicolumn{1}{c}{Averaging} & \multicolumn{1}{c}{Ranking} \\
&& \multicolumn{1}{c}{$q_{i1}'$} & \multicolumn{1}{c}{${\rm pos}_1(i)$} & \multicolumn{1}{c}{$q_{i2}'$} & \multicolumn{1}{c}{${\rm pos}_2(i)$} & \multicolumn{1}{c}{$q_{i3}'$} & \multicolumn{1}{c}{${\rm pos}_3(i)$} & & \multicolumn{1}{c}{$\sum_{j=1}^3 q_{ij}'/N$} & \multicolumn{1}{c}{$\sum_{j=1}^3{n-{\rm pos}_{j}(i)}$}  \\
 \vspace{0.2cm}
& Project 1 & $7.1$ & 1 & $-11.7$ & 3 & $4.4$ & 1 & & $-0.07$ & 4 \\ 
& Project 2 & $2.0$ & 3 & $2.0$ & 1 & $2.0$ & 2 & & $2.00$ & 3 \\
& Project 3 & $5.5$ & 2 & $-4.1$ & 2 & $-1.8$ & 3 & & $-0.13$ & 2 \\
\end{tabular}
\vspace{2mm}
				\begin{tablenotes}[para,flushleft]
				\SingleSpacedXI\footnotesize
					\textit{Notes.} The data are from a sample realization with $n=3$ projects of type $t_1=10$, $t_2=5$,and $t_3=0$, respectively. Knowledge breadth of $\beta=0$ has all $N=3$ agents endowed with expertise value $e_{\rm M}=5$. True project qualities are $q_1=3$, $q_2=2$, and $q_3=1$. The quantity $q_{ij}'$ denotes the quality of project $i\in\{1,2,3\}$ as perceived by agent $j\in\{1,2,3\}$ and ${\rm pos}_j(i)$ is the position of project $i$ in the preference list of agent $j$. In the case of $m=1$ selected projects, the Averaging rule would select Project 2 whereas Ranking would select Project 1.
				\end{tablenotes}
			\end{threeparttable}
		\end{adjustbox}
	\end{center}
\end{table}

If the organization instead used Averaging for the same data, it would select Project \#2 first, as it receives a mean agent assessment of $2$. Project \#1 would be second-most attractive, with a mean assessment of $-0.07$. Project \#3 would be least attractive, with a mean assessment of $-0.13$. The aggregate organizational preference list produced by Averaging would not list the best project first, because it is vulnerable to a single agent's misclassification.

\begin{figure}[htbp]
\FIGURE
    {\includegraphics[width=.8\textwidth]{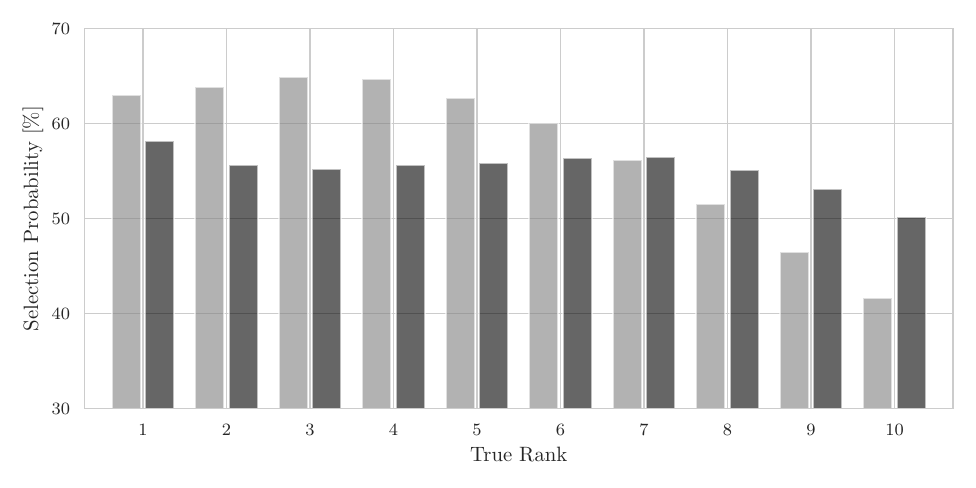}}
    {Classification Effectiveness\label{fig:misclassification}}
    {The figure charts the probability that Ranking (\tikz\node[gray!80,fill=gray!80,rectangle,draw] {};) and Averaging (\tikz\node[black!60,fill=black!60,rectangle,draw] {};) select a project with true quality rank $\{1,\dots,m\}$ into a portfolio of $m=10$ projects when considering $n=100$ project candidates. Knowledge breadth here is $\beta=0$, and the type and quality distributions are $\mathcal{U}(0,10)$ and $\mathcal{U}(-5,5)$, respectively. Probabilities are derived as the relative selection frequencies in $2\times 10^5$ i.i.d.\ realizations.}
\end{figure}

In our base model with $n=100$, $m=10$, $\beta=0$, and $N=3$, Ranking identifies the highest-quality project in about 63\% of $2\times 10^5$ realizations, whereas Averaging does so in about 58\% of cases (see the selection probabilities in Figure~\ref{fig:misclassification}). The reason is that agents with an outlying impression of project quality can sway the aggregate selection more readily in the Averaging protocol than in the Ranking protocol. Ranking accommodates extreme inputs more readily, because uncapped quality differences are translated into capped score differences in rank orders (the maximum rank-order score difference is $n-1$ per agent). For the Ranking protocol to misclassify a project in aggregate, a relatively greater number of individual agents would have to concurrently misclassify. 

Ranking is thus particularly effective in identifying projects of extreme quality. This ability to discriminate is crucial for portfolio selection with tight budgets. Discrimination effectiveness is less relevant for larger budgets: If a 15th-best project is misclassified as 17th best, for instance, the impact on portfolio performance is marginal. Consequently, Averaging gains in relative performance when budgets also permit the selection of more moderate-quality projects. The flatter project-selection probability distribution of Averaging is more suited to more munificent budgets. Selecting more projects balances the impact of misclassifications. As $m$ approaches $n$, the performance of all selection protocols becomes equal.

The performance dynamic of Ranking and Averaging resembles that observed for sample mean and sample median as gauges of population values. For normal distributions, the sample mean is more efficient than the sample median in estimating the mean value of the underlying population [\ie, the variance of the sample mean is smaller than that of the sample median \citep{kenney1962relation}]. However, the sample median is known to be less sensitive to small-sample outliers that can introduce unwanted bias in the sample mean. In accordance with these arguments, we show in Section~\ref{sec:crowds_vs_experts} that Averaging achieves higher portfolio performance than Ranking for a large number of agents $N$. Most organizational selection committees, however, consist of only a small number of decision makers. For them, Ranking is a more effective aggregation rule than Averaging.
\subsection{Budgets and Choice Sets}
\label{sec:budgets}

The size of an organization's innovation budget determines the number of projects $m$ it can select. And the number of project alternatives $n$ available and identified by an organization compose the choice set from which it can select. While the former typically pales in comparison to the size of the latter \citep{klingebiel2018risk}, numbers can vary across organizations. Such variance could matter in principle, since values for $m$ and $n$ bound the theoretically attainable portfolio performance. The supplemental analyses reported in Appendix \ref{app:sensitivity}, however, shows that they rarely change the relative ordering of aggregation-rule performance as observed in our base-case analysis.

\begin{figure}[htbp]
    \FIGURE
    {\includegraphics{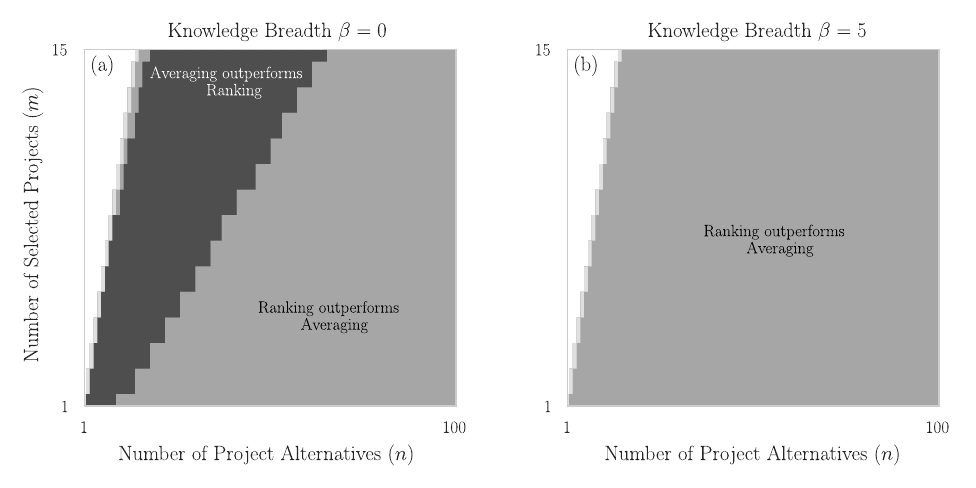}}
    {Aggregation-Rule Dominance\label{fig:ranking_vs_averaging}}
    {The panels show regions in $(m,n)$-space where either Averaging (\tikz\node[black,fill=black, rectangle, minimum width=10pt,draw] {};) or Ranking (\tikz\node[gray!80,fill=gray!80, rectangle, minimum width=10pt,draw] {};) outperforms the other. Note that the performance of Averaging is equal to that of Ranking for $m=n$. The project-quality distribution is $\mathcal{U}(-5,5)$; that of project types is $\mathcal{U}(0,10)$. Three agents have expertise values $e_1\equiv e_{\rm M} = 5$, $e_2=e_{\rm M}-\beta$, and $e_3=e_{\rm M}+\beta$, respectively. All results are based on ensemble means for $2\times 10^5$ i.i.d.\ realizations.}
\end{figure}

An interesting edge case is that of small choice sets. When the cardinality of the choice set of candidate projects in our model is smaller than about $n=10$, and $N=3$ agents with knowledge breadth $\beta=0$ were to select $m=1$ project, Averaging outperforms Ranking (see Figure~\ref{fig:ranking_vs_averaging}a), as it has access to more information on the underlying project quality. For larger numbers of candidate projects and small values of $m$, Averaging is more likely to misclassify a project. 

Another edge case to Ranking's dominance is found for generous budgets and zero knowledge breadth. When the number of projects $m$ that the budget permits is not much smaller than the number of projects $n$ available in the choice set, Averaging outperforms Ranking for $\beta=0$ (see Figure~\ref{fig:ranking_vs_averaging}a). In such cases, the benefit of greater information provision outweighs the low risk of misclassification. This effect is restricted to organizations with homogeneous decision makers. 

If the knowledge breadth $\beta$ takes on more realistic values above zero, such that available expertise values are better aligned with the underlying project types, the advantage of Averaging over Ranking diminishes. Figure~\ref{fig:ranking_vs_averaging}b illustrates that with $\beta=5$, Ranking outperforms Averaging even for the smallest possible choice set with $n=2$ elements. Further simulations indicate that this dominance begins at even lower knowledge breadths \textemdash~results for $\beta=2.5$ are consistent with those obtained for $\beta=5$.
\subsection{Delegation Errors}
\label{sec:delegation_error}
Innovation projects contain novel elements for which past data offers limited guidance. Experts from some domains will have relevant experience and, through associations, may gauge the promise of novelty better than other experts. But organizations may not always know ex ante who these most suitable experts are, leading to errors in delegation. Such likelihood of delegation error is one reason for why academic journals, as well as grant institutions \citep{bian2022innofund}, for example, seek the opinion of multiple expert reviewers without fully delegating decisions to any.

In Figure~\ref{fig:selection_rules}, we show the selection performance of delegating to project experts as a function of delegation error $r\in\{0,0.5,1\}$. When $r = 0$, projects are always assigned to the most qualified expert, whereas with $r = 1$, projects are randomly distributed among the three available agents. In more mathematical terms, organizations assign projects with probability $r/3$ to any of the two least suitable agents and with probability $1-2r/3$ to the most suitable ~\citep{csaszar2013organizational}. 

Detailed simulations for a larger number of values of $r$ show that for $r\gtrsim 0.2$ the Delegation protocol no longer provides a substantially better performance than Ranking for a small budget (see Figure~\ref{fig:selection_rules}a). The influence of delegation errors diminishes with larger budgets (see Figure~\ref{fig:selection_rules}b). An error of $r=0.2$ means that 87\% of projects are evaluated by an appropriate expert. 

Although ascertaining delegation-error rates in prior empirical work is limited by the lack of counterfactuals, it is not hard to imagine that innovation projects, covering novel terrain by definition, are often mismatched to expertise in existing terrain. Ambiguity about the suitability of experts in evaluating innovation thus renders delegation an unattractive aggregation rule.

In an alternative approach, organizations could try delegating project evaluation to a single Portfolio Expert, whose expertise minimizes the uncertainty with respect to all projects. In our main specification, this would be the agent with expertise $e_{\rm M}=5$, which is equal to the mean project type. A Portfolio Expert would thus perform as well as an Individual. Erroneously designating as portfolio expert one of the other agents with expertise values $e_{\rm M}\pm \beta$ would yield a performance that is worse than that of the Individual protocol for $\beta>0$. 
\subsection{Environmental Turbulence}
\label{sec:type_variation}

The performance of different selection rules does not depend only on the level of knowledge breadth in the group of decision makers but also on the distribution and range of project types. When market environments shift, the relevance of organizations' existing knowledge base diminishes. \cite{csaszar2013organizational} exemplify such shifts with the technological transition from analog to digital photography, which rendered some of Polaroid's expertise less useful for project selection \citep{tripsas2000polaroid}. Considering additional type distributions helps us examine how different aggregation rules cope with environmental shifts.

\begin{figure}[htbp]
    \FIGURE
    {\includegraphics[width=\textwidth]{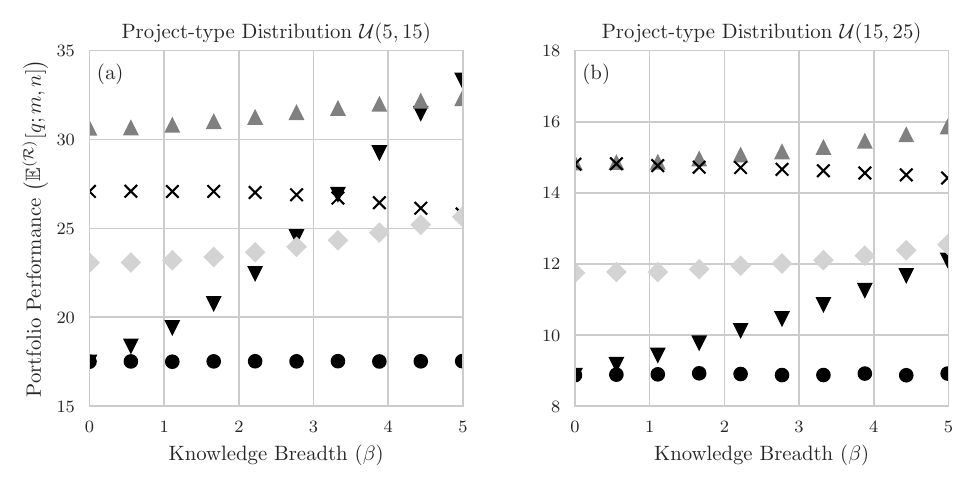}}
    {Shifting Project Landscapes\label{fig:selection_rules_type_distr}}
    {The panels show the total performance of project portfolios selected by the aggregation rules of Individual (\tikz\node[black,fill=black,circle,inner sep=2pt,draw] {};), Delegation (\tikz\node[black,fill=black,isosceles triangle,isosceles triangle apex angle=60,rotate=270,inner sep=1.8pt,draw] {};) with $r=0$, Voting (\tikz\node[lightgray,fill=lightgray,diamond,inner sep=2pt,draw] {};), Averaging (\tikz\node[black,fill=black,cross out,inner sep=2pt,line width=1pt,draw] {};), and Ranking (\tikz\node[gray!80,fill=gray!80,isosceles triangle,isosceles triangle apex angle=60,rotate=90,inner sep=1.8pt,draw] {};). All rules select $m=10$ projects from a choice set of $n=100$ candidates with quality distribution $\mathcal{U}(-5,5)$. The expertise of Individual agents is $e_{\rm M}=5$. In all other rules, we consider $N=3$ agents and the expertise value of agent $j\in\{1,\dots,N\}$ is $e_j=e_{\rm M}-\beta+\frac{2\beta}{N-1} (j-1)$. The project-type distribution $\mathcal{U}(5,15)$ of Panel (a) overlaps with the expertise of fewer agents than the base-case distribution $\mathcal{U}(0,10)$ reported in Figure \ref{fig:selection_rules}. Project-type distribution $\mathcal{U}(15,25)$ of Panel (b) has no overlap. All results are based on ensemble means of $2\times 10^5$ i.i.d.\ realizations.}
\end{figure}

Figure~\ref{fig:selection_rules_type_distr} reports the portfolio performance of aggregation rules for the type distributions $\mathcal{U}(5,15)$ and $\mathcal{U}(15,25)$. Performance generally decreases when the distance between required and available expertise increases. If the expertise of decision makers is close to the type of the project under evaluation, selection errors are small. Consequently, an expertise level of $e_{\rm M}=5$ yields smaller errors for the type distribution $\mathcal{U}(0,10)$ than $\mathcal{U}(5,15)$, for example. 

Ranking, however, is relatively less impacted by risk of misclassification when project-type distributions shift. The Ranking rule's performance surpasses that of error-free Delegation for knowledge breadth as wide as $\beta \gtrsim 2$ for the base-case type distribution $\mathcal{U}(0,10)$, and as wide as $\beta \gtrsim 5$ for the type distribution $\mathcal{U}(5,15)$. The further the project-type distribution moves from agents with relevant expertise, the greater the knowledge breadth at which Ranking outperforms even perfect Delegation. Relatively homogeneous organizations facing disruptive change would thus fare best with Ranking.
\subsection{Crowds versus Experts}
\label{sec:crowds_vs_experts}
Up to this point, we kept the number of decision makers at a constant $N=3$. Relaxing this constraint can reveal relative differences in the marginal benefit of additional decision makers. Increasing crowd size also allows collectives to outperform experts even in settings where delegation error is absent and expertise broadly distributed \citep{davis2014crowd}. 

Through approaches such as open innovation or open strategy \citep{chesbrough2006open,stadler2021open} organizations can enlarge their pool of internal decision makers, and it would be instructive to know how large such collectives would need to be to outperform delegation to three knowledgeable project experts. IBM, a large technology firm with an in-house crowd effort, managed to have $25$ colleagues review projects of its iFundIT program, though not everyone evaluated all projects \citep{feldmann2014ibm}. We could take this observation as an upper bound of the number of suitable agents that organizations might feasibly recruit to the collective task of portfolio selection.\footnote{Open-science initiatives may worry less about innovation appropriation \citep{altman2022ecosystems,arora2016openip} and could thus attract larger numbers of assessors from outside the organization than IBM managed from within. EteRNA (\url{https://eternagame.org}), for example, enlists outsiders to select the most promising molecule designs for resource-intensive testing. Governments are another type of organization that could tap a greater pool of decision makers for selecting projects in participative-budgeting exercises, such as through the Consul project (\url{https://consulproject.org})} 

\begin{figure}[htbp]
    \FIGURE
    {\includegraphics[width=\textwidth]{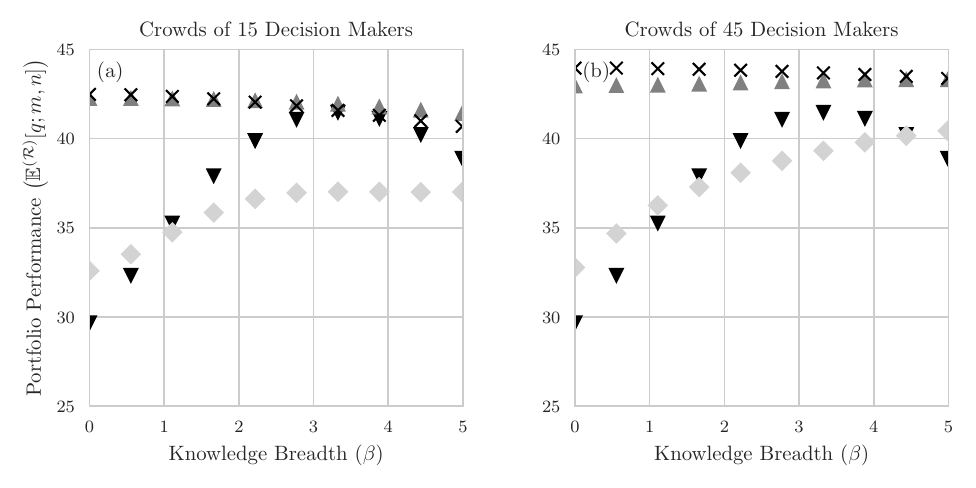}}
    {Crowds versus Three Experts\label{fig:selection_rules_number_voters}}
   {The panels display the portfolio performance of crowd-aggregation rules Voting (\tikz\node[lightgray,fill=lightgray,diamond,inner sep=2pt,draw] {};), Averaging (\tikz\node[black,fill=black,cross out,inner sep=2pt,line width=1pt,draw] {};), and Ranking (\tikz\node[gray!80,fill=gray!80,isosceles triangle,isosceles triangle apex angle=60,rotate=90,inner sep=1.8pt,draw] {};), alongside the performance achievable with error-free Delegation (\tikz\node[black,fill=black,isosceles triangle,isosceles triangle apex angle=60,rotate=270,inner sep=1.8pt,draw] {};) to three in-house experts. The number of crowd decision makers change \textemdash~all other parameters remain the same as in the base case considered earlier. As the number of decision makers increases, all crowd-aggregation rules eventually outperform Delegation to the three project experts. All results are based on ensemble means of $2\times 10^5$ i.i.d.\ realizations.}
\end{figure}

We thus examine the number of decision makers required for collective protocols to outperform Delegation to the three project experts of our base-case parameterization (Figure~\ref{fig:selection_rules_number_voters} illustrates crowds of $N=15$ and $N=45$). Averaging outperforms Delegation to project experts as the number of decision makers $N$ nears $15$; Ranking already does at around $N=13$. Voting can compete with Delegation over the whole range of knowledge breadth only with $45$ or more decision makers. 

While Ranking outperforms Averaging with about ten or fewer decision makers, the order reverses with bigger crowds (see Figure~\ref{fig:selection_rules_number_voters}), even at large values of knowledge breadth. In simulations with $N=100,200,500,1000$, this performance gap grows (2.57\%, 2.86\%, 3.06\%, and 3.12\%, respectively, at $\beta=0$). The magnitude of the growing gap might nonetheless be insufficient to justify the use of Averaging, given that such large crowds would be hard to manage and well in excess of those observed as feasible in the IBM study of \cite{feldmann2014ibm}.

In all studied scenarios, Voting is inferior to Averaging and Ranking. In particular, for $\beta=0$, the performance of Voting changes only very little with an increase of the crowd size, even if it is by an order of magnitude. This is because in Voting, agents make binary choices, where all projects perceived to yield positive payoffs receive approval. When noise is within bounds and expertise overlaps, there is limited benefit to soliciting more near-identical decisions from a crowd. 
Ranking and Averaging gain more from homogeneous crowds as they provide more fine-grained information for selection. 

In a converse scenario with considerable noise and/or knowledge breadth, Voting (very) slowly gains in performance with an increasing number of decision makers. Each additional decision maker adds granularity to the aggregation scale (three decision makers mean that a project can have either no, one, two, or three votes \textemdash~ten decision makers would classify a project anywhere between no and ten votes, and so on). Ranking and Averaging provide granular aggregation scales even with few decision makers.
\subsection{Batching}
\label{sec:cognition}
In the aggregation protocols we study, agents evaluate each project on its own. One could alternatively imagine agents directly comparing projects and making relative judgments \textemdash~at least when there is no strict need to first provide separate assessments, such as with Voting or Ranking. In such cases, cognitive limitations might weaken comparison effectiveness as the number of candidate projects grows. At some level of $n$, agent evaluations may become unreliable.

To guard against such scenario, one could design an evaluation regime in which individuals receive no more projects than they are able to compare reliably. The precise magnitude of such a cognitive limit $c$ is unknown and varies with context \citep{scheibehenne2010overload}\footnote{The members of the Academy of Motion Pictures and Sciences, for example, rank between five and ten candidates to collectively select the Best Picture \citep{economist2015oscars}. In the lab, participants predicting league tables appear able to rank 30-odd sports teams without apparent difficulty \citep{lee2014thurston}. Other lab participants appeared to struggle with the comparisons necessary for the ranking of eight Kickstarter project candidates \citep{cui2019scoring}.}. The illustrative analysis reported below sets the limit to a conservative batch size of $c=10$. The idea is that, when an organization's choice set is as large as the $n=100$ projects considered in our base-case analyses, agents could share the load and each evaluate $c=10$ projects only. 

Reducing agents' cognitive load requires proportionally more of them. The number of agents in our base-case analyses would have to go up by a factor of $n/c$ to ensure that each project gets the same number of evaluations in the cognition-conscious batching regime. 

If little is known ex-ante about projects and agents, agents will receive a randomly drawn subset of $c$ projects. The evaluation could also be shared among an organization's cohort of evaluators on the basis of preference \citep{bentert2020batching}. A more directed approach is to allocate $c$ projects each to $N$ agents such that there is a match between the types of expertise required and available. The organization would ask its relatively most experienced colleagues to vote, estimate, or rank\footnote{Practical examples of delegating a subset of candidates to assessors on the basis of perceived expertise include, for example, the selection process for the Academy of Management's Technology and Innovation Management division Best Dissertation award shortlist. Documented in the literature is the selection of treatments via ranking by groups of orthodontists \cite[see][]{li2022bayesian} before making the final selection. Moods of Norway had employees rank products of a category with which they are familiar to estimate future demand for apparel \citep{salikhov2021forecasting}. Geographically separate juries also select through ranking the semi-finalists for the Eurovision Song Contest. Each jury accepts a quota \citep{ginsburgh2023eurovision}.}. This makes most sense when evaluators and projects are known to span a comparable range of expertise. 

Finally, the organization may authorize those subgroups of evaluators to make decisions on its behalf. Innovating organizations often acknowledge limits to the comparability of projects of different departments, subdividing the overall budget and allowing departments to make their own decisions about which projects to select \citep{chao2008buckets}.

\begin{figure}
    \FIGURE
    {\includegraphics[width=\textwidth]{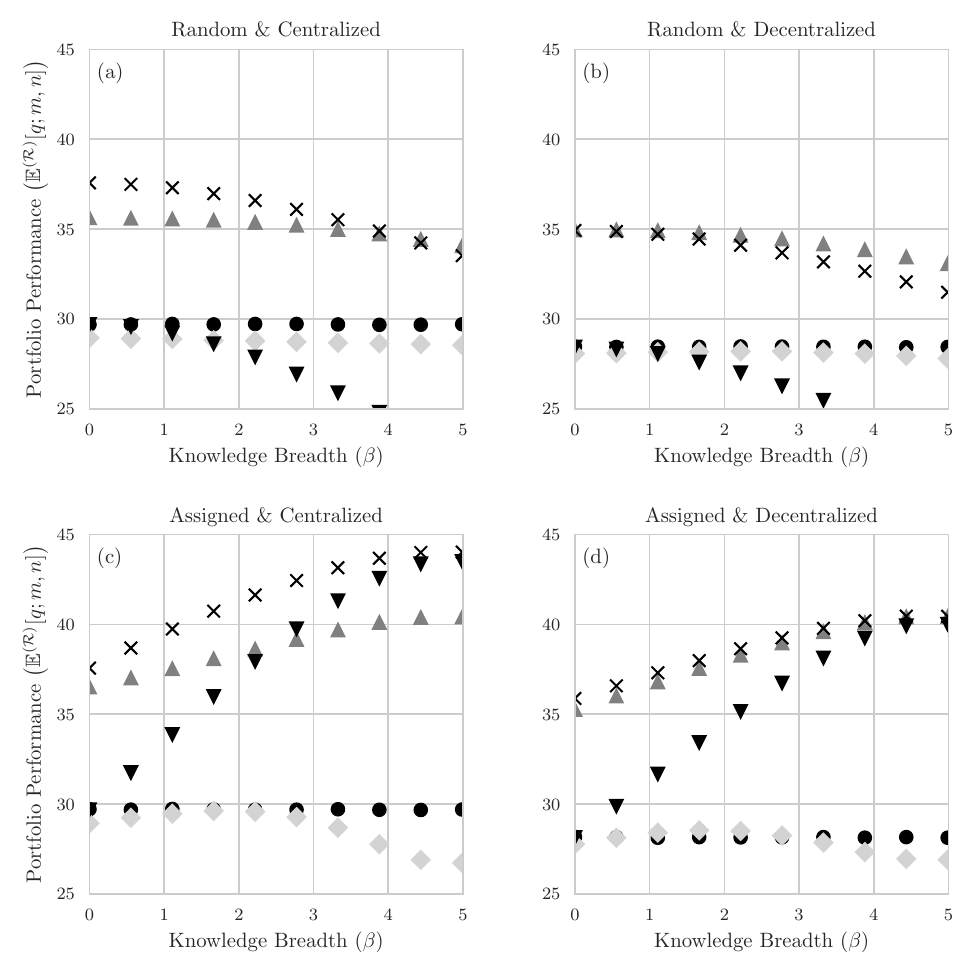}}
    {Batched Portfolio Selection\label{fig:batching}}
    {The panels show the total performance of project portfolios selected by the aggregation rules of Individual (\tikz\node[black,fill=black,circle,inner sep=2pt,draw] {};), Delegation (\tikz\node[black,fill=black,isosceles triangle,isosceles triangle apex angle=60,rotate=270,inner sep=1.8pt,draw] {};), Voting (\tikz\node[lightgray,fill=lightgray,diamond,inner sep=2pt,draw] {};), Averaging (\tikz\node[black,fill=black,cross out,inner sep=2pt,line width=1pt,draw] {};), and Ranking (\tikz\node[gray!80,fill=gray!80,isosceles triangle,isosceles triangle apex angle=60,rotate=90,inner sep=1.8pt,draw] {};). The organization selects $m=10$ projects from a total of $n=100$ available projects. Each agent receives a batch of $c=10$ projects for consideration. To cover all projects, the Individual and Delegation rules involve $N=10$ agents. The collective selection rules involve $N=30$. The expertise of Individual agents is $e_{\rm M}=5$. In all other rules, the expertise value of agent $j\in\{1,\dots,N\}$ is $e_j=e_{\rm M}-\beta+\frac{2\beta}{N-1} (j-1)$. In the models used to generate the data in the two top panels, projects are distributed uniformly at random among agents, whereas those employed in the two bottom panels allocate projects based on agent expertise. The project-type and quality distributions are $\mathcal{U}(0,10)$ and $\mathcal{U}(-5,5)$, respectively. All results are based on ensemble means of $2\times 10^5$ i.i.d.\ realizations. The theoretical maximum for the expected quality associated with selecting $m$ out of $n$ available projects [see Eq.~\eqref{eq:uniform_e_max}] is approximately 44.6.}
\end{figure}

Figure~\ref{fig:batching} reports the analysis for these approaches to batching. As the main analyses, batching is based on uniform type, quality, and expertise distributions, maintaining the number of project candidates at $n=100$ and the number of selected projects at $m=10$. We multiply the number of agents involved in each selection rule by $n/c$, yielding $N=10$ for Individual and Delegation, $N=30$ for Voting, Averaging, and Ranking.\footnote{If the value of $n/c$ is not an integer, it is rounded to the nearest integer. The same goes for $m/c$.} Each agent receives a batch of $c=10$ projects to asses. 

Without expertise matching, the assignment of $c=10$ projects is uniformly at random without replacement from the pool of $n=100$ projects. Expertise matching is a hard problem and a thorough review of the multitude of implementation possibilities goes beyond the scope of our work. We here employ simple ordinal matching. We begin by arranging projects in ascending type order and agents in ascending expertise order. We then assign the first batch of $c=10$ projects to the first agent, in the case of Individual and Delegation, or the first three agents, in the case of Voting, Averaging, and Ranking. The second batch goes to the second agent(s), and so on.

Agents normally submit their project votes, estimates, or ranks to a central organization for the final aggregate selection decision. In a decentralized setting, by contrast, each of $n/c$ agents, or sets of agents, selects $m/c$ projects. In the analysis of Figure~\ref{fig:batching}, this means one project each. Collectively, these $m$ selected projects make up an organization's portfolio.

The results reported in Figure~\ref{fig:batching} show that the performance of Averaging improves relative to Ranking, at least at lower levels of knowledge breadth $\beta$. This is because aggregating ten project ranks from three agents yields less granular distinctions than aggregating precise project estimates. Although agents' detailed project estimates may be flawed, the random tiebreakers often necessary in aggregating rankings are relatively more detrimental to portfolio selection. Therefore, if cognitive limitations are a concern, evaluation noise moderate, and agents plentiful, Averaging may offer a more effective batch-selection method than Ranking.

The results reported in Figure~\ref{fig:batching} also show that random batching unsurprisingly underperforms expertise batching, especially when knowledge breadth $\beta$ increases. Real-world organizations will find themselves somewhere in between the random and perfect expertise assignment. 

Decentralizing decision rights, too, is usually a bad idea, due to the loss of being able to optimize at the portfolio rather than sub-portfolio level. Ranking, however, suffers less from decentralization than other rules. This is because the projects that would have been selected at the sub-portfolio level also often end up being selected at the portfolio level. The top projects of each batch also have the top scores in the portfolio. It is rare that the second-placed project in one batch has a greater sum of inverted ranks than the first-placed of another batch. Therefore, if cognitive limitations were a concern and addressed with batching, organizations that use a Ranking rule could more easily decentralize with less of a performance sacrifice. 
\section{Discussion}
We extend earlier work on aggregating project approval to the context of selecting projects for resource-limited portfolios. We show that Ranking, an aggregation process specific to portfolio selection, is often more effective than Averaging and Voting, processes also available in a single-project approval context. These findings contribute to the literatures on resource allocation and aggregation, respectively.
\subsection{Resource Allocation Decisions}
The earlier work of \cite{csaszar2013organizational} highlights how the choice of rules for aggregating individual decisions into an organizational one can produce meaningful performance differences. Its insights are applicable to contexts in which the (dis)approval of one project is viewed independent of the (dis)approval of other projects (see the assumptions in \cite{sharapovdahlander2021, malenko2023vc, piezunka2023, criscuolo2017evaluating}, for example). Also relevant for the isolated approval of projects are attempts to aggregate project forecasts into decisions through polls or markets~\citep{atanasov2017distilling,lamberson2012forecasts}.

Acknowledging, however, that organizations are resource-constrained, means that not all projects that would be approvable in isolation can be funded. The challenge for organizations is to identify the subset of many possible projects that most likely maximize organizations' return on investment \citep{archer2007project,kleinmuntz_2007,sharpe1998smithkline}. Solving such optimization problems involves preference orders, derived from aggregating individual agent preferences. 

In this portfolio-selection context, the relative performance differences among aggregation rules reported by \cite{csaszar2013organizational} do not hold. While the earlier study is justifiably concerned with the performance of all approved projects, the focus for portfolio selection is on the performance of only those projects that organizations can afford to fund. That is because resource allocation in organizations is not only about correctly identifying projects with positive returns but about selecting the subset of projects that deliver the greatest return on the investable resources \citep{brasil2019product, klingebiel2018risk}.

Our work reveals how totaling project ranks provided by agents offers the highest aggregation-rule performance in many circumstances that one might find in organizations. The ranking-rule performance is below the optimum that omniscient decision makers could attain, but it is above the performance of other rules for aggregating decisions with limited information. 

By highlighting performance dynamics of decision aggregation rules, our work provides a normative foundation for descriptive research on resource allocation. Crucially, it provides a baseline benchmark for work attempting to highlight behavioral inefficiencies in portfolio selection (\cite{criscuolo2017evaluating, sommer2020search}, for example). It also provides a reference point without which empirical observations of portfolio-selection rule performance \citep{sharapovdahlander2021, malenko2023vc} are hard to interpret. 

In future research, it would be valuable to expand upon our work by considering additional factors such as differential project types and costs \citep{goel2019knapsack} or dynamic features \citep{si2022managing}. Further opportunities arise from merging our insights with those on managing portfolios under uncertainty, including the partial allocation and potential re-allocation of resources over time \citep{klingebiel2021optionality}, the allocation of resources by more than one organization \citep{folta1998governance}, or the incentive structures used to populate choice sets for portfolio selection \citep{klingebiel2022motivating}.
\subsection{Organizational Decision Aggregation}
Our work further contributes to the resurgent interest in aggregation structures \citep{christensen2021context, keum2017influence,bottcher2021tradeoffs}. In particular, we shed further light on situations in which one might expect expert decision makers to outperform variously aggregated crowds \citep{csaszar2018individual, scott2020experts, mannes2014selectcrowds}. Specifically, choosing the best subset from a range of options non-trivially departs from previously studied contexts due to its greater need for discrimination.

Although delegation performs highest in settings where experts can be found, the often imperfect organizational process of matching uncertain projects with the right domain specialists in turbulent environments calls for alternative approaches. Having multiple imperfectly informed decision makers weigh in on the same project propositions typically improves on the eventual performance that an organization can expect from its portfolio. Ranking does so most effectively.

When agents rank projects, they provide an assessment of how the quality of one project compares to that of others. Most rankings in real-world organizations are necessarily imperfect amalgamations of multiple criteria, ranging from profit forecasts over strategic fit to short-term versus long-term considerations\footnote{In the project and portfolio literatures, rankings already feature heavily: They are outputs of organizational prioritization efforts \citep{kornish2017research,schilling2023book}. Our work underlines that rankings also have a place as inputs to those efforts.}. Using subjective rankings as an input to the ultimate organizational decision thus makes intuitive sense. In contrast to seemingly more precise project-value appraisals, crude rankings often help select higher-performing project portfolios.

Future field research on aggregating agents' preference lists may benefit from the fact that Ranking endogenizes a concern over strategic behavior. Employees who know of their organization's resource constraints may not provide project assessments or votes that reflect their true beliefs, in an attempt to lift some projects above the cut-off \citep{bassi2015voting}. With Ranking, agents maximize the chances of their organization funding their favorite projects by ranking projects in the preferred order. There is little room for gaming by submitting preference orders that fail to reflect beliefs (unlike with Averaging, for example, where agents could inflate forecasts for their preferred projects, and deflate those for less preferred candidates). 

Similarly beneficial is that Ranking appears more tolerant of biased inputs, requiring fewer agents to select optimal sets than alternative aggregation methods \citep{boehmer2023bias}. Ranking methods that additionally reflect preference intensity \citep{skowron2020participatory} might be similarly robust to strategy and bias, presenting a straightforward extension possibility for our current work.

Further opportunities for future research include the extension of our work by accounting for quadratic voting effects~\citep{eguia2019quadratic}. An alternative direction to consider involves devising algorithms that can help identify effective selection rules akin to algorithmic solutions of multi-winner election problems~\citep{peters2018single,xia2022groupai}. Future work may also explore project-cost distributions \citep{benade2021preference}, skill heterogeneity and weighted aggregation \citep{benyashar2021skill, manouchehrabadi2022democratic}, strategic and coordinated selection behavior~\citep{myatt2007theory}, as well as vote trading~\citep{casella2021does}. Further potential exists in recognizing the impact of organizational competition, which may favor contrarian rules such as minority voting \citep{arrieta2023minority,malenko2023vc}.
\subsection{Managerial Application}

\begin{figure}[htbp]
    \FIGURE
    {\includegraphics[width=0.7\textwidth]{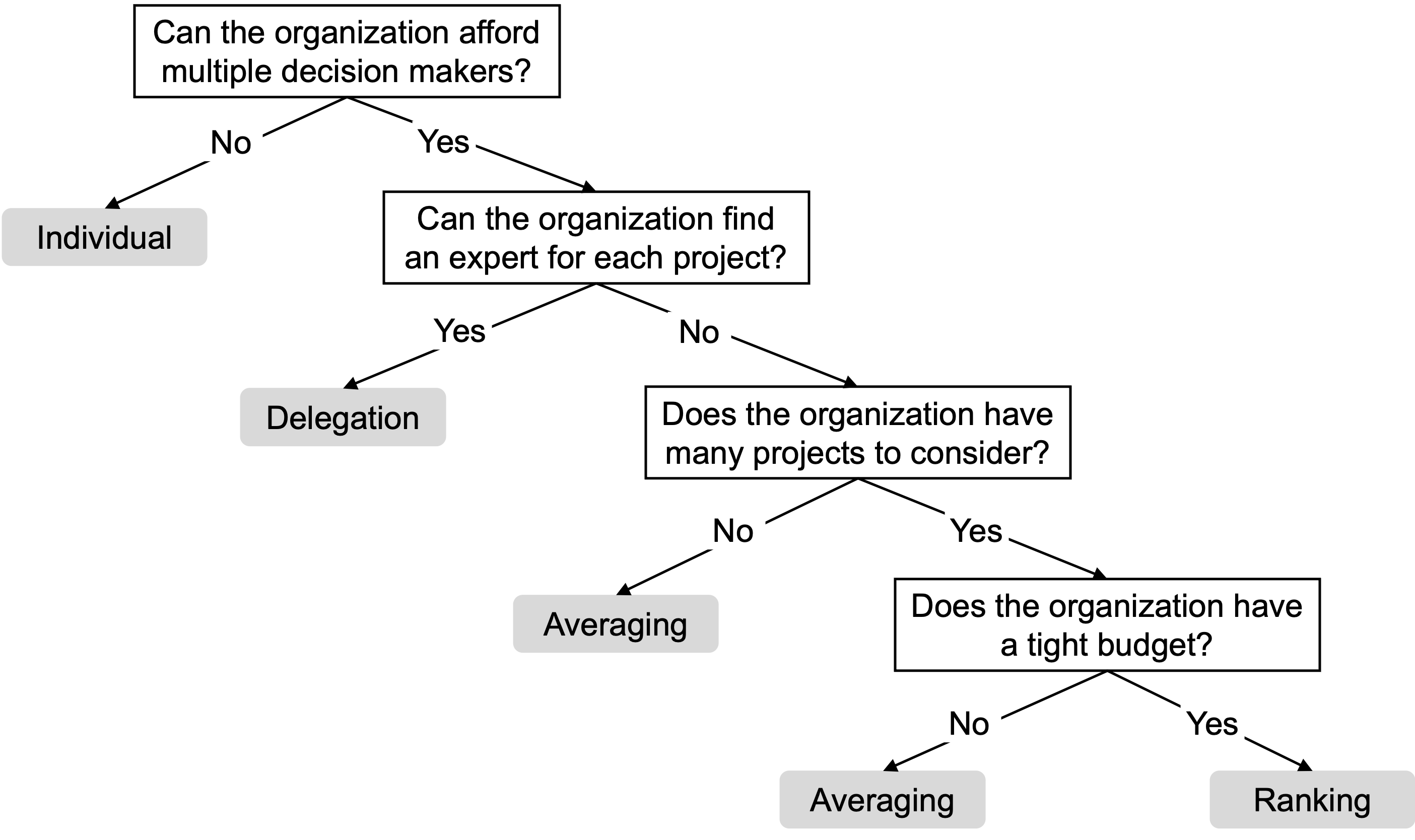}}
    {Choosing a Selection Method\label{fig:decision_tree}\vspace{3mm}}
    {When asking more than one person to decide on which projects to select for an innovation portfolio, organizations stand to benefit from adopting a Ranking rule. Averaging is a good choice in specific situations. Simple Voting is less suited to the selection of innovation projects.}
\end{figure}

The performance of aggregation rules depends on the availability of information about the knowledge held by employees and the size of the innovation budget and choice set. Organizations looking for simplified guidance on which rule to adopt may consider the illustrative decision tree presented in Figure~\ref{fig:decision_tree}. In portfolio-selection situations with many choices, tight budgets, and unclear expertise, our work recommends the Ranking rule.

The performance of the Ranking protocol is good news for two reasons. One is that many organizations already informally aggregate rankings in some form when they meet in a committee setting. Such committee meetings often involve discussions that contribute to the convergence of individuals' assessments of projects \citep{lane2022herding}. The Ranking protocol deals well with low belief heterogeneity, attenuating the potential impact of convergent beliefs. Therefore, organizational reality may not be too far from feasible aggregation optima.

The second reason is that organizations probably have an easier time implementing a Ranking protocol than some of the other aggregation mechanisms reviewed here. Rather than having to submit seemingly precise project-value assessments, decisions makers simply have to put projects into a preference order. This may become more taxing as the number of projects to consider increases, but aggregation through ranking is somewhat forgiving of the accuracy of assessments that lead to the preference orders. It often produces innovation portfolios with the relatively highest performance outcomes. 

Given these advantages, what could go wrong? A few aspects of the Ranking rule's practical application might be worth paying attention to in future empirical research. A first step would be studies of safeguards against loss in judgement quality that stems from the greater cognitive load of comparing a potentially large number of candidates simultaneously \citep{cui2019scoring,gelauff2024rank}. Our main models sidestep this issue by having agents score projects individually, which only later amounts to a ranked project list for each agent (the aforementioned procedure for the Aggregate Ranking of Top Universities does the same). Innovating organizations might get close to this ideal by having project proposals presented one at a time, making comparisons easier to avoid \citep{mussweiler2009compare}. 

To then further mitigate potential order effects, whereby evaluators compare a focal proposal to what they can remember about those evaluated previously \citep[e.g.][]{elhorst2021contest,klingebiel2022sample}, organizations might wish to shuffle the sequence of proposals for each evaluator. The setting of a committee meeting does not easily lend itself to different evaluation sequences but asynchronous online assessments would. One challenge for such asynchronous assessments would be to ensure that assessors evaluate all candidates. Incomplete rankings akin to those submitted on participatory budgeting platforms such as Stanford's\footnote{\url{https://pbstanford.org}}, for example \textemdash~where assessors receive no compensation and thus prioritize attention \textemdash~not only provide less information (as per Section \ref{sec:cognition}) but also open the door to herding and influencing.

Moreover, future research could examine the effectiveness with which organizations are able to aggregate the rankings that their employees provide. Without an explicit aggregation rule, managers' processing of rank information may differ from their processing of scores. For example, \cite{chun2022ranks} suggest that people sometimes treat rankings as a shortcut heuristic for separating top candidates from a cohort, forfeiting more fine-grained discrimination. Automating aggregation may thus prove useful in guarding against processing biases. In any case, adding Ranking to the list of aggregation methods to be examined behaviorally \citep{niederberger2023scales} seems apt given its conceptual benefits for innovation-portfolio selection.
\subsection{Conclusion}
Our work contributes to the understanding of resource allocation in innovation portfolios. Increasing data availability and scholarly interest in the topic have revealed interesting patterns of behavior when multiple organizational actors make joint decisions. Yet, interpreting their relevance requires a normative foundation. In providing one, we show that some insights, such as about the effects of knowledge breadth and delegation error, apply in the context of portfolio project selection decision just as they do in the better-known but less applicable context of isolated project approvals. However, portfolio selection additionally requires discrimination between projects and the relative performance ordering of suitable decision-aggregation rules thus changes. 

Our results indicate that Ranking is the most effective selection rule, especially in unstable market environments, and often outperforms Averaging even for small values of knowledge breadth. In many scenarios, ranking is preferable to other aggregation rules. Delegation makes sense when companies can assign each project to a relevant expert. But environmental turbulence can cause Ranking to outperform even perfect Delegation. 

Multi-candidate selection may be relevant not only in the context of innovation, but also for other organizational decisions under uncertainty \citep{klingebiel2023ambiguity}, including investments in personnel or technology. Our work thus contributes to a better understanding of selection regimes within organizations. The choice of an appropriate aggregation rule is a discretionary element in the design of resource allocation processes that has substantial performance implications.
\section*{Code Availability}
Our source codes are publicly available at \url{https://gitlab.com/ComputationalScience/multiwinner-selection}.
\ACKNOWLEDGMENT{Lucas Böttcher acknowledges financial support from hessian.AI and the Army Research Office (grant W911NF-23-1-0129)} 
\bibliographystyle{informs2014}
\bibliography{refs.bib}
\clearpage

\begin{APPENDICES}
\section{Performance Sensitivity to Budget and Choice Set}
\label{app:sensitivity}
\smallskip
\subsection{Theoretical Performance Limits}
\label{app:performance_properties}
For $m\in\{1,\dots,n\}$ selected projects, we use $\mathbb{E}^*[q;m,n]$ to denote the theoretical performance maximum. It can be derived from the order statistic~\citep{david2004order} of the underlying project quality distribution.\footnote{Order statistics have also been employed by \cite{einhorn1977quality} to mathematically characterize an aggregation rule in which $N$ individuals with varying levels of expertise evaluate a single project ($n=1$).} For $n$ realizations of the random variable $q_i \sim \phi$, one obtains the order statistic $q_{(i)}$ by sorting the realizations $q_i$ in ascending order. The value of $\mathbb{E}^*[q;m,n]$ is found by evaluating
\begin{equation}
\mathbb{E}^*[q;m,n]=\sum_{i=n+1-m}^n \int_{\underline{q}}^{\overline{q}} q\, f_{q_{(i)}}(q)\,\mathrm{d}q\,,
\label{eq:max_performance}
\end{equation}
where $f_{q_{(i)}}(q)$ denotes the probability density function (PDF) of the order statistic $q_{(i)}$ with support $[\underline{q},\overline{q}]$. Dividing $\mathbb{E}^*[q;m,n]$ by the number of selected projects $m$, yields the expected theoretical quality maximum per selected project $\mathbb{E}_m^*[q;m,n]=\mathbb{E}^*[q;m,n]/m$. 

The PDF of the order statistic $q_{(i)}$ in Eq.~\eqref{eq:max_performance} is given by
\begin{equation}
f_{q_{(i)}}(q) = \frac{n!}{(i-1)!(n-i)!} \phi(q) [ \Phi(q) ]^{i-1} [ 1 - \Phi(q) ]^{n-i}\,,
\label{eq:pdf_order_statistic}
\end{equation}
where $\Phi(x)$ is the CDF of the project quality distribution $\phi(x)$. Using the transformation ${u=\Phi(q)}$, the PDF of the quantity $u_{(i)}$ is a beta distribution~\citep{gentle2019computational} with shape parameters $i$ and $n+1-i$. That is,

\begin{equation}
f_{u_{(i)}}(u) = \frac{n!}{(i-1)!(n-i)!} u^{i-1} [ 1 - u ]^{n-i}\,.
\end{equation}

The mean of the beta distribution $f_{u_{(i)}}(u)$ is $i/(n+1)$, so we can compute the theoretical performance maximum for any uniform distribution $\mathcal{U}(\underline{q},\overline{q})$ according to

\begin{equation}
\mathbb{E}^*[q;m,n] = m\underline{q} +(\overline{q}-\underline{q})\sum_{i=n+1-m}^n\frac{i}{n+1}=m \left[\underline{q} +(\overline{q}-\underline{q})\frac{2n+1-m}{2n+2}\right]\,.
\label{eq:uniform_e_max_2}
\end{equation}

In the limit of a large number of candidate projects, the proportion of selected projects, or selectiveness \citep{klingebiel2021optionality}, at which $\mathbb{E}^*[q;m,n]$ reaches its peak value is\footnote{Real-world organizations cannot confidently gauge the shape of the distribution of payoffs from the innovation projects proposed to them, and they consequently determine the size of their budget more pragmatically \citep[cf.][]{sengul2019allocation,stein2003agency}. Additionally, Appendix \ref{app:distributions} shows how portfolio performance depends on the shape of the underlying quality distribution.} 
\begin{equation}
 m^*/n=
    \begin{cases}
       1/(1-\underline{q}/\overline{q})\, & \text{if } \underline{q}\leq 0,\overline{q}>0\\
       1\, & \text{if } \underline{q}> 0,\overline{q}>0\\\
       0\, & \text{if } \underline{q}< 0,\overline{q}<0\,.
    \end{cases}
    \label{eq:performance_peak}
\end{equation}
In the same limit, the maximum expected quality per selected project $\mathbb{E}_m^*[q;m,n]=\mathbb{E}^*[q;m,n]/m$ can be approximated by $\mathbb{E}_m^*[q;m,n]\approx \underline{q}+(\overline{q}-\underline{q})[1-m/(2n)]$ [see Eq.~\eqref{eq:uniform_e_max_2}]. As the selectiveness $m/n$ approaches 1, the quantity $\mathbb{E}_m^*[q;m,n]$ approaches $\underline{q}+(\overline{q}-\underline{q})/2$ for a uniform quality distribution $\mathcal{U}(\underline{q},\overline{q})$. Also notice that $\mathbb{E}_m^*[q;m,n]$ approaches the upper limit of the underlying uniform quality distribution, $\bar{q}$, as $m/n$ approaches 0. 

According to Eq.~\eqref{eq:uniform_e_max_2}, for any uniform quality distribution $\mathcal{U}(\underline{q},\overline{q})$, the performance measures $\mathbb{E}^*[q;m,n]$ and $\mathbb{E}_m^*[q;m,n]=\mathbb{E}^*[q;m,n]/m$ depend on both the number of projects $m$ that an organization's budget permits to select and the total number of projects $n$ in the choice set. For sufficiently large numbers of projects, we have $\mathbb{E}_m^*[q;m,n]\approx \underline{q}+(\overline{q}-\underline{q})[1-m/(2n)]$.

Figure~\ref{fig:max_quality}a shows $\mathbb{E}_m^*[q;m,n]$ as a function of $m$ for different values of $n$ and for a uniform quality distribution $\mathcal{U}(-5,5)$. The largest value of $\mathbb{E}_m^*[q;m,n]$ for constant $n$ is $5(n-1)/(n+1)$, and it is obtained for $m=1$. For $n=20,50,100$, the corresponding values are about 4.5, 4.8, and 4.9, respectively. As $m$ approaches $n$, the maximum performance $\mathbb{E}_m^*[q;m,n]$ approaches 0 (see Figure~\ref{fig:max_quality}a).

\begin{figure}[htbp]
    \FIGURE
    {\includegraphics{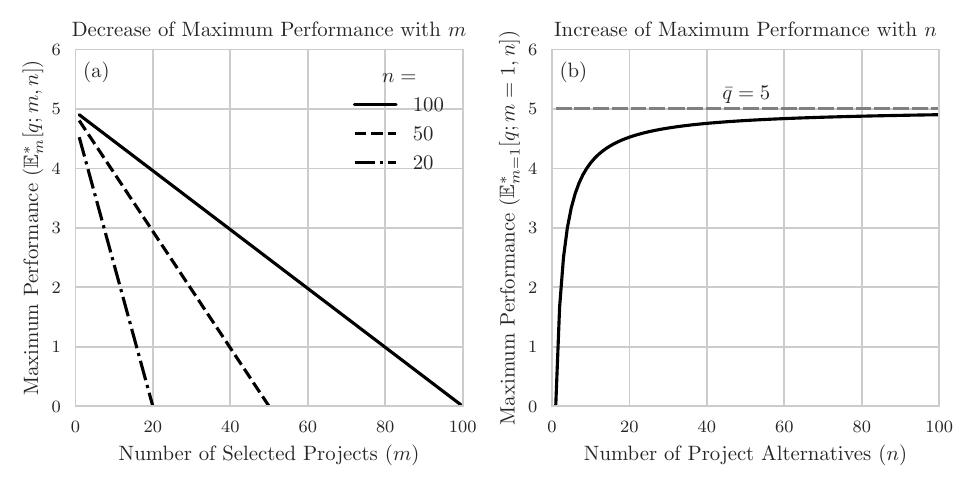}}
    {Theoretical Maxima.\label{fig:max_quality}}
    {Panel (a) charts the theoretical maximum of the expected quality per selected project, $\mathbb{E}_m^*[q;m,n]$, as a function of the number of projects $m\leq n$ permitted by an organization's budget, for choice sets with different numbers of project alternatives $n$. Panel (b) charts the theoretical maximum performance of a single selected project, $\mathbb{E}_{m=1}^*[q;m=1,n]$. With the uniform quality distribution $\phi=\mathcal{U}(-5,5)$ considered in our base-case analysis, the performance measure approaches $\bar{q}=5$ in the limit of $n\rightarrow\infty$.}
\end{figure}

To visualize the dependence of $\mathbb{E}_m^*[q;m,n]$ on $n$ for constant $m$, we show in Figure~\ref{fig:max_quality}b the performance measure $\mathbb{E}_{m=1}^*[q;m=1,n]$ for a single selected project as a function of $n$. The quality distribution is again $\mathcal{U}(-5,5)$. We observe that an increase in the number of available projects from 0 to 10 is associated with a large increase in $\mathbb{E}_{m=1}^*[q;m=1,n]$ from 0 to more than 4. Increasing $n$ from 10 to 100 yields a much smaller increase in $\mathbb{E}_{m=1}^*[q;m=1,n]$ of about 0.8. In the limit $n\rightarrow\infty$, the maximum performance $\mathbb{E}_{m=1}^*[q;m=1,n]$ approaches $\bar{q}=5$. Although more available projects yield a larger value of $\mathbb{E}_{m}^*[q;m,n]$ for a given $m$, possible performance gains that are associated with further increasing $n$ may be negligible if $m/n\ll 1$.

\begin{figure}[htbp]
    \FIGURE
    {\includegraphics{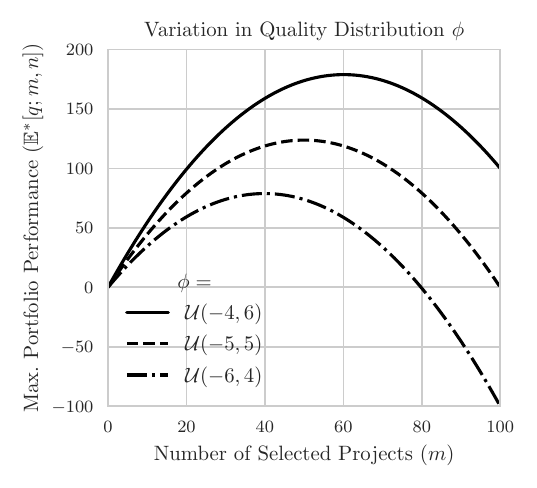}}
    {Maximum Portfolio Performance for Varying Quality Distributions\label{fig:max_portfolio_quality}}
    {The figure charts the theoretical maximum of the expected portfolio performance, $\mathbb{E}^*[q;m,n]$, as a function of the number of projects $m\leq n$ permitted by an organization's budget, for quality distributions $\phi$ with different support. The number of available project candidates is $n=100$.}
\end{figure}

In addition to achieving a high performance per selected project, one often wishes to optimize the overall portfolio performance, whose theoretical maximum is $\mathbb{E}^*[q;m,n]=m\mathbb{E}_m^*[q;m,n]$. For a uniform quality distribution $\mathcal{U}(\underline{q},\overline{q})$, we have $\mathbb{E}^*[q;m,n]=m\{\underline{q}+(\overline{q}-\underline{q})[1-m/(2n)]\}
$ [see Eq.~\eqref{eq:uniform_e_max_2}]. Figure~\ref{fig:max_portfolio_quality} shows $\mathbb{E}^*[q;m,n]$ as a function of $m$ for three different uniform quality distributions. The optimum of $\mathbb{E}^*[q;m,n]$ is attained for 
\begin{equation}
    m^*=\frac{\underline{q}+\overline{q}+2\overline{q}n}{2(\overline{q}-\underline{q})}\,.
    \label{eq:m_star}
\end{equation}
For the quality distributions used in Figure~\ref{fig:max_portfolio_quality}, the corresponding rounded values of $m^*$ are $40$, $50$, and $60$. Using Eq.~\eqref{eq:m_star}, the optimal selectiveness $m^*/n$ approaches $1/(1-\underline{q}/\overline{q})$ in the limit of large $n$. Given the constraint $0 \leq m^*/n \leq 1$ for the optimal selectiveness, we obtain Eq.~\eqref{eq:performance_peak} in the large-$n$ limit.
\smallskip
\subsection{Relative Performance Ordering}
\label{app:relative_ordering}
The maximum performance per selected project, $\mathbb{E}_m^*[q;m,n]$, provides an upper bound for the aggregation-rule performance $\mathbb{E}_m^{(\mathcal{R})}[q;m,n]$. Figure~\ref{fig:max_quality_comparison}a,b shows that the $(m,n)$-dependence of $\mathbb{E}_m^{(\mathcal{R})}[q;m,n]$ associated with different aggregation rules is similar to the $(m,n)$-dependence of $\mathbb{E}_m^*[q;m,n]$. 

The performance measure $\mathbb{E}^{(\mathcal{R})}[q;m,n]$ exhibits a pronounced initial increase with $n$, gradually diminishing in magnitude for larger values of $n$ (see Figure~\ref{fig:max_quality_comparison}c,d). In accordance with the results presented in the main text (Section~\ref{sec:base_case}), Ranking and Averaging perform well for a small knowledge breadth (see Figure~\ref{fig:max_quality_comparison}a,c) while Delegation (without delegation errors) is closer to the maximum performance for large values of knowledge breadth (see Figure~\ref{fig:max_quality_comparison}b,d). The relative performance ordering of aggregation rules is consistent with the results reported in the main text.
\begin{figure}[htbp]
    \FIGURE
    {\includegraphics{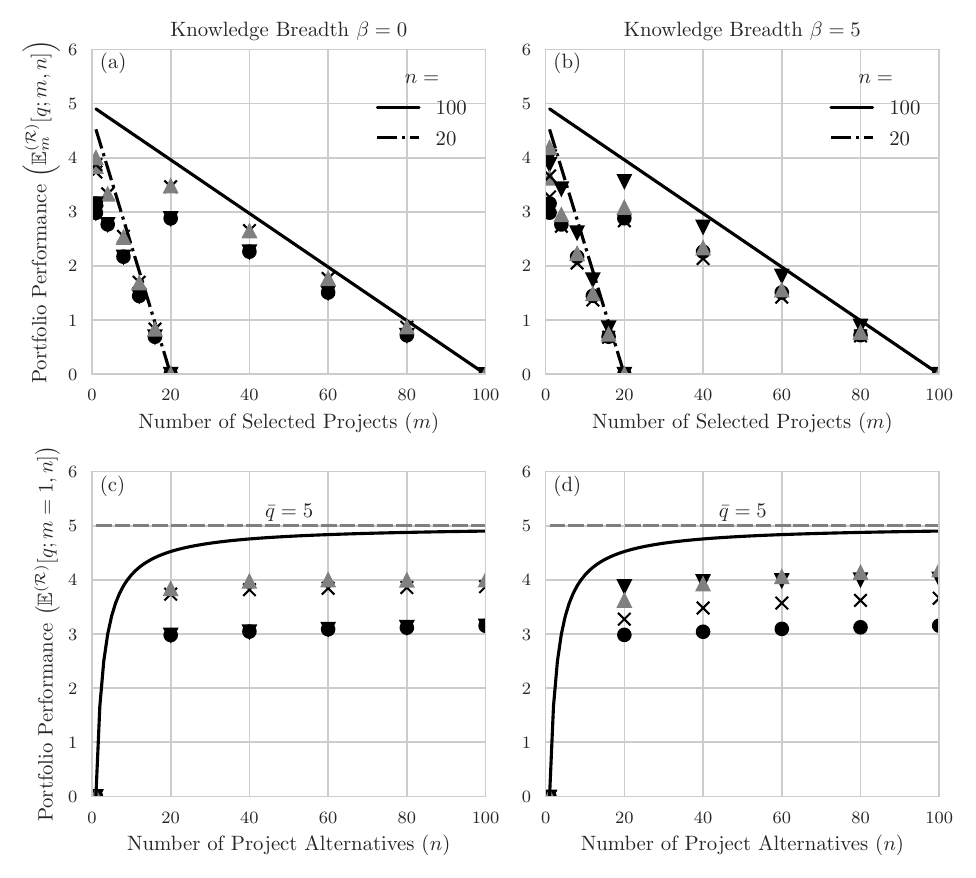}}
    {Budget and Choice-Set Sensitivity\label{fig:max_quality_comparison}}
    {The panels show the performance per selected project and portfolio performance associated with the aggregation rules Individual (\tikz\node[black,fill=black,circle,inner sep=2pt,draw] {};), Delegation (\tikz\node[black,fill=black,isosceles triangle,isosceles triangle apex angle=60,rotate=270,inner sep=1.8pt,draw] {};) with $r=0$, Ranking (\tikz\node[gray!80,fill=gray!80,isosceles triangle,isosceles triangle apex angle=60,rotate=90,inner sep=1.8pt,draw] {};), and Averaging (\tikz\node[black,fill=black,cross out,inner sep=2pt,line width=1pt,draw] {};). The quality and type distributions are $\mathcal{U}(-5,5)$ and $\mathcal{U}(0,10)$, respectively. The expertise value of the agent under the Individual rule is $e_{\rm M}=5$. All other rules use three agents with expertise values $e_{1}=e_{\rm M}$, $e_2=e_{\rm M}-\beta$, and $e_3=e_{\rm M}+\beta$. We set $\beta=0$ and $5$ in panels (a,c) and (b,d), respectively. The solid and dash-dotted black lines in panels (a,b) indicate the theoretical maximum portfolio performance per selected project, $\mathbb{E}_m^*[q;m,n]$, for $n=100$ and $20$, respectively. The solid black line in panels (c,d) is is the theoretical maximum portfolio performance of a single selected project, $\mathbb{E}_{m=1}^*[q;m=1,n]$. It approaches $\bar{q}=5$ (dashed grey line) in the limit $n\rightarrow\infty$. All results are based on ensemble means of $2\times 10^5$ i.i.d.\ realizations.}
\end{figure}
\clearpage
\subsection{Relative Performance with Very Small Budgets}
In Section~\ref{sec:base_case}, we studied the portfolio performance of different aggregation rules for $m=10,30$. In Figure~\ref{fig:selection_rules_extended}, we compare the portfolio performance of all the aggregation rules considered for smaller budgets with $m=1,3$. The relative positioning of the aggregation rules in these two cases aligns with the case where $m=10$. However, for $m=1$ and intermediate knowledge breadths, there are no discernible performance differences between Ranking and error-free Delegation.
\begin{figure}[htbp]
    \FIGURE
    {\includegraphics[width=\textwidth]{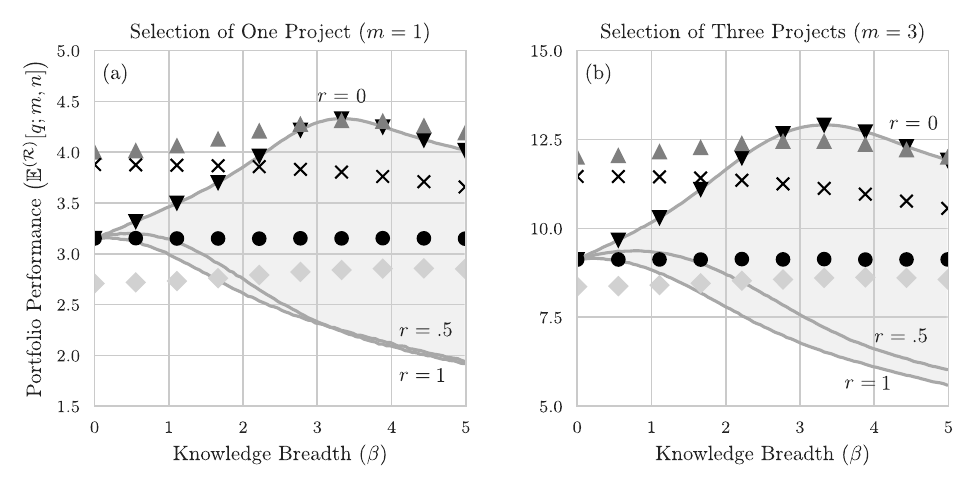}}
    {Very Small Budgets\label{fig:selection_rules_extended}}
    {The panels show the total performance of project portfolios selected by the aggregation rules of Individual \tikz\node[black,fill=black,circle,inner sep=2pt,draw] {};), Delegation (\tikz\node[black,fill=black,isosceles triangle,isosceles triangle apex angle=60,rotate=270,inner sep=1.8pt,draw] {};), Voting (\tikz\node[lightgray,fill=lightgray,diamond,inner sep=2pt,draw] {};), Averaging (\tikz\node[black,fill=black,cross out,inner sep=2pt,line width=1pt,draw] {};), and Ranking (\tikz\node[gray!80,fill=gray!80,isosceles triangle,isosceles triangle apex angle=60,rotate=90,inner sep=1.8pt,draw] {};). The shaded area delineates the performance range for the Delegation rule as determined by the delegation error $r$ defined in Section~\ref{sec:delegation_error}. The results are derived from subsets of $c=30$ preselected projects out of a total of $n=100$ available projects. The budget is $m=1$ in Panel (a) and $m=3$ in Panel (b). The expertise value of the agent under the Individual rule is $e_{\rm M}=5$. All other rules use three agents with expertise values $e_1=e_{\rm M}$, $e_2=e_{\rm M}-\beta$, and $e_3=e_{\rm M}+\beta$. The project-type and quality distributions are $\mathcal{U}(0,10)$ and $\mathcal{U}(-5,5)$, respectively. All results are based on ensemble means of $4\times 10^5$ i.i.d.\ realizations. The theoretical maxima for the expected quality [see Eq.~\eqref{eq:uniform_e_max}] are approximately 4.9 ($m=1$) and 14.4 ($m=3$).}
\end{figure}
\clearpage
\section{Performance Sensitivity to Project Distributions}
\label{app:distributions}
In addition to the uniform quality distribution $\phi=\mathcal{U}(\underline{q},\overline{q})$ discussed in the main text, we explore the impact of variations in the quality distribution on portfolio performance. We examine two additional quality distributions: (i) a truncated normal distribution $\widetilde{\mathcal{N}}(0,1,\underline{q},\overline{q})$ with a mean of zero and unit variance, and (ii) a power-law distribution with an exponent of $-1/2$.

Both additional distributions have support $[\underline{q},\overline{q}]$, and for our analysis, we set $\underline{q}=-5$ and $\overline{q}=5$, consistent with the base-case analysis in the main text.

In contrast to a uniform quality type distribution where projects occur with equal probability, regardless of their quality, the truncated normal distribution leads to fewer occurrences of projects with large negative or positive qualities. Projects with qualities close to zero have higher probabilities of occurrence in this distribution.

Regarding the power-law distribution we consider, on average, approximately 70\% of the projects will have negative quality. Moreover, only about 5\% of the project qualities will exceed a value of 4. This distribution represents scenarios where only a relatively small number of projects are associated with relatively large positive qualities.

For the two quality distributions under consideration, Figure~\ref{fig:quality_trunc_normal} charts the performance of aggregation rules for $n=100$ projects and $m=10,30$ selected projects as a function of knowledge breadth. Both distributions encompass fewer high-quality projects than the uniform distribution analyzed in the main text, resulting in a lower overall portfolio performance. The shown differences in portfolio performance are consistent with the findings reported in the main text. 

\begin{figure}[htbp]
    \FIGURE
    {\includegraphics[width=\textwidth]{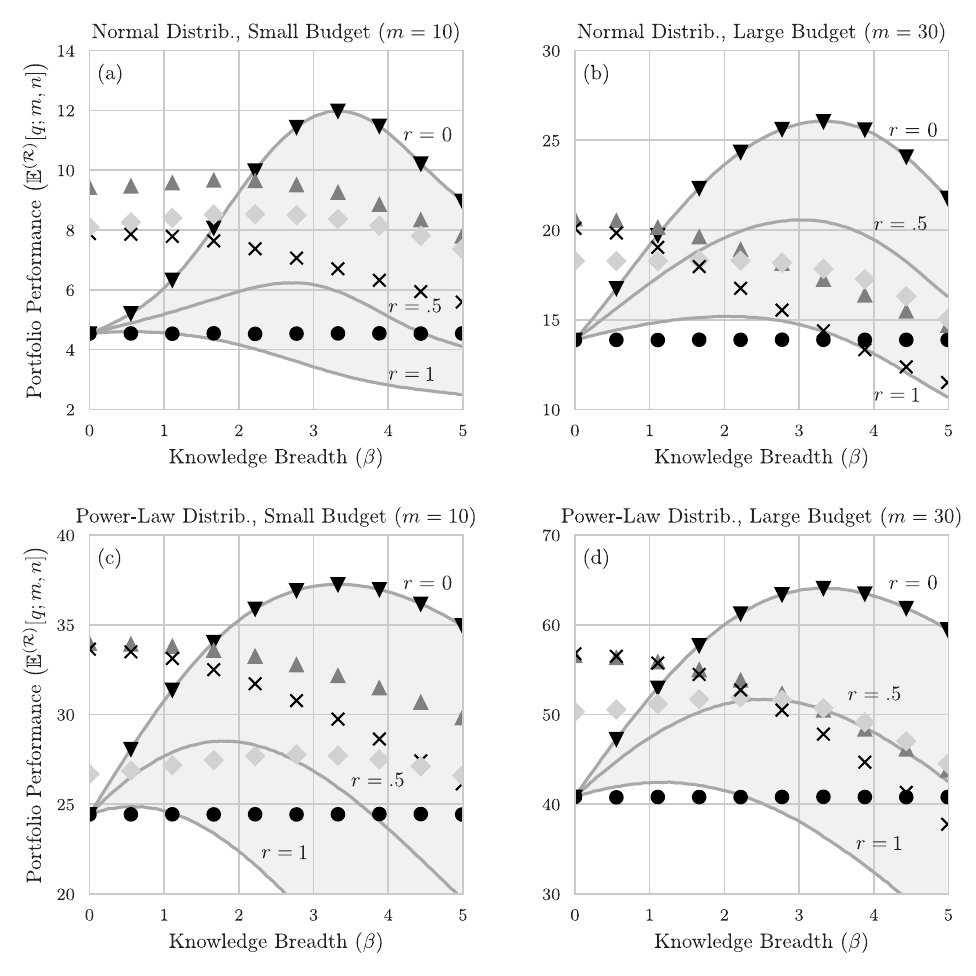}}
    {Alternative Project-Quality Distributions\label{fig:quality_trunc_normal}}
    {The panels show the total performance of project portfolios selected by the aggregation rules of Individual (\tikz\node[black,fill=black,circle,inner sep=2pt,draw] {};), Delegation (\tikz\node[black,fill=black,isosceles triangle,isosceles triangle apex angle=60,rotate=270,inner sep=1.8pt,draw] {};), Voting (\tikz\node[lightgray,fill=lightgray,diamond,inner sep=2pt,draw] {};), Averaging (\tikz\node[black,fill=black,cross out,inner sep=2pt,line width=1pt,draw] {};), and Ranking (\tikz\node[gray!80,fill=gray!80,isosceles triangle,isosceles triangle apex angle=60,rotate=90,inner sep=1.8pt,draw] {};). The shaded area delineates the performance range for the Delegation rule as determined by the delegation-error parameter $r$ defined in Section~\ref{sec:delegation_error}. Results are based on $n=100$ available projects. The numbers of selected projects is $m=10$ in panels (a,c) and $m=30$ in panels (b,d). The expertise value of the agent under the Individual rule is $e_{\rm M}=5$. All other rules use three agents with expertise values $e_{1}=e_{\rm M}$, $e_2=e_{\rm M}-\beta$, and $e_3=e_{\rm M}+\beta$. The project-type distribution is $\mathcal{U}(0,10)$. In Panels (a) and (b), we use a truncated normal distribution $\widetilde{\mathcal{N}}(0,1,-5,5)$ with zero mean and unit variance as quality distribution. In Panels (c) and (d), the quality distribution is a power-law distribution with an exponent of $-1/2$ and a support of $[-5,5]$. All results are based on ensemble means of $4\times 10^5$ i.i.d.\ realizations. The theoretical performance maximum [see Eq.~\eqref{eq:uniform_e_max}] is approximately 17.3 in Panel (a), 34.5 in Panel (b), 39.5 in Panel (c), and 67.5 in Panel (d).}
\end{figure}

Voting performs substantially better in the simulations with truncated normal distributions centered on zero than in simulations with uniform project-quality distributions. This is because of the many projects with near-zero quality. Whereas uniform distributions favor decision rules that detect relative quality differences between projects, narrow zero-centered normal distributions predominantly require detection of whether or not a project has a positive value. Voting's coarseness more easily achieves the latter. The effectiveness of Voting in portfolio selection from normally distributed projects thus comes to resemble its effectiveness in approving uniformly distributed projects in isolation \citep{csaszar2013organizational}.
\end{APPENDICES}
\end{document}